%% file: m_charm_nf3.tex
\documentclass[11pt,a4paper]{article}
\pdfoutput=1   
\usepackage[utf8]{inputenc}
\usepackage{tabularx}
\usepackage{mathtools}
\usepackage{longtable}

\usepackage{alpha}
\hypersetup{%
   pdftitle    = {Determination of the charm quark mass in lattice QCD with 2+1 flavours on fine lattices},
   pdfauthor   = {J. Heitger, F. Joswig, S. Kuberski},
   pdfkeywords = {Lattice QCD, Nonperturbative effects, Quark masses, Charm}
}%
\usepackage{lmodern}
\usepackage[outercaption]{sidecap}
\usepackage{xspace}
\usepackage{cancel}
\usepackage{caption}
\usepackage{subcaption}
\usepackage{array}
\usepackage{lscape}
\usepackage{microtype}
\usepackage{rotating}
\usepackage{scalerel}
\usepackage{tikz}
\usetikzlibrary{svg.path}

\definecolor{orcidlogocol}{HTML}{A6CE39}
\tikzset{
	orcidlogo/.pic={
		\fill[orcidlogocol] svg{M256,128c0,70.7-57.3,128-128,128C57.3,256,0,198.7,0,128C0,57.3,57.3,0,128,0C198.7,0,256,57.3,256,128z};
		\fill[white] svg{M86.3,186.2H70.9V79.1h15.4v48.4V186.2z}
		svg{M108.9,79.1h41.6c39.6,0,57,28.3,57,53.6c0,27.5-21.5,53.6-56.8,53.6h-41.8V79.1z M124.3,172.4h24.5c34.9,0,42.9-26.5,42.9-39.7c0-21.5-13.7-39.7-43.7-39.7h-23.7V172.4z}
		svg{M88.7,56.8c0,5.5-4.5,10.1-10.1,10.1c-5.6,0-10.1-4.6-10.1-10.1c0-5.6,4.5-10.1,10.1-10.1C84.2,46.7,88.7,51.3,88.7,56.8z};
	}
}

\newcommand\orcidicon[1]{\href{https://orcid.org/#1}{\mbox{\scalerel*{
				\begin{tikzpicture}[yscale=-1,transform shape]
				\pic{orcidlogo};
				\end{tikzpicture}
			}{|}}}}

\newcommand\blfootnote[1]{%
	\begingroup
	\renewcommand\thefootnote{}\footnote{#1}%
	\addtocounter{footnote}{-1}%
	\endgroup
}

\begin{document}

\input{title.tex}

\makeatletter
\g@addto@macro\bfseries{\boldmath}
\makeatother

\input{01_introduction}
\input{02_theory}

\input{03_computations}
\input{04_extrapolations}
\input{05_results}
\input{06_conclusions}

\begin{acknowledgement}%
\input{acknow.tex}
\end{acknowledgement}

\newpage
\appendix
\input{app_quark_mass_running}
\clearpage
\input{app_tables}

\newpage
\small
\addcontentsline{toc}{section}{References}
\bibliographystyle{JHEP}
\bibliography{library}

\end{document}

%% file: title.tex
\preprintno{%
MS-TP-21-01\\
\vfill
}

\title{%
Determination of the charm quark mass in lattice QCD with $2+1$ flavours on fine lattices
}

\collaboration{\includegraphics[width=2.8cm]{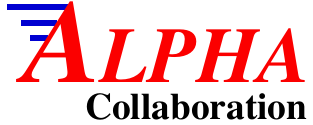}}

\author{Jochen Heitger\,\orcidicon{0000-0001-5181-9145}, Fabian Joswig\,\orcidicon{0000-0003-0740-6524} and Simon Kuberski\,\orcidicon{0000-0002-0955-9228}$^\dagger$\blfootnote{${}^\dagger$Present address: \textit{Helmholtz-Institut Mainz, Johannes Gutenberg-Universit\"at Mainz, 55099 Mainz, Germany}}}

\address{Westf\"alische Wilhelms-Universit\"at M\"unster, Institut f\"ur Theoretische Physik,\\
Wilhelm-Klemm-Stra{\ss}e 9, 48149 M\"unster, Germany}

\begin{abstract}
  We present a determination of the charm quark mass in lattice QCD with three active quark flavours. The calculation is based on PCAC masses extracted from $N_\mathrm{f}=2+1$ flavour gauge field ensembles at five different lattice spacings in a range from 0.087\,fm down to 0.039\,fm. The lattice action consists of the $\mathrm{O}(a)$ improved Wilson-clover action and a tree-level improved Symanzik gauge action. Quark masses are non-perturbatively $\mathrm{O}(a)$ improved employing the Symanzik-counterterms available for this discretisation of QCD. To relate the bare mass at a specified low-energy scale with the renormalisation group invariant mass in the continuum limit, we use the non-pertubatively known factors that account for the running of the quark masses as well as for their renormalisation at hadronic scales. We obtain the renormalisation group invariant charm quark mass at the physical point of the three-flavour theory to be $M_\mathrm{c} = 1486(21)\,$MeV. Combining this result with five-loop perturbation theory and the corresponding decoupling relations in the $\overline{\mathrm{MS}}$ scheme, one arrives at a result for the renormalisation group invariant charm quark mass in the four-flavour theory of $M_\mathrm{c}(N_\mathrm{f}=4) = 1548(23)\,$MeV, where effects associated with the absence of a charmed, sea quark in the non-perturbative evaluation of the QCD path integral are not accounted for. In the $\overline{\mathrm{MS}}$ scheme, and at finite energy scales conventional in phenomenology, we quote $m^{\overline{\mathrm{MS}}}_{\mathrm{c}}(m^{\overline{\mathrm{MS}}}_{\mathrm{c}}; N_\mathrm{f}=4)=1296(19)\,\mathrm{MeV}$ and $m^{\overline{\mathrm{MS}}}_{\mathrm{c}}(3\,\mathrm{GeV}; N_\mathrm{f}=4)=1007(16)\,\mathrm{MeV}$ for the renormalised charm quark mass.\\
\end{abstract}

\begin{keyword}
Lattice QCD \sep Non-perturbative effects \sep Quark masses  \sep Charm%
\end{keyword}

\maketitle

%% file: 01_introduction.tex
\section{Introduction}
Apart from the strong coupling constant, quark masses are the only fundamental parameters of quantum chromodynamics (QCD), the theory of the strong interaction. As such, their precise knowledge is not only of general interest, but also essential in the search for new physics beyond the Standard Model of particle physics. In particular, heavy quark masses serve as key parametric inputs for tests of the Standard Model, via hadronic contributions by QCD matrix elements to weak decays and ensuing constraints on CKM matrix elements \cite{Antonelli:2009ws}, as well as in Higgs branching ratios to charm and bottom quarks; see for instance Refs.~\cite{Heinemeyer:2013tqa,Petrov:2015jea}.
The extraction of these and other Standard Model parameters that have their origin in the strong interaction requires to evaluate the QCD path integral. Since quarks are confined in bound states and cannot propagate freely, their properties have to be investigated in the hadronic regime at low energies where perturbative methods fail. We base our calculations on lattice QCD, where the theory is formulated on a discretised space-time grid to enable, in conjunction with Monte Carlo simulation techniques, an ab initio and thereby non-perturbative solution of the underlying highly non-linear equations with, in principle, full control of statistical uncertainties and systematic effects.

Our determination of the charm quark mass is performed on a subset of the gauge configuration ensembles, generated by the {\it Coordinated Lattice Simulations (CLS)} cooperative effort of European lattice QCD teams, with $N_\mathrm{f}=2+1$ flavours of non-perturbatively $\mathrm{O}(a)$ improved Wilson quarks and the tree-level Symanzik-improved gauge action \cite{Bruno:2014jqa,Mohler:2017wnb}. The sea quark content in these discretised QCD configurations comprises a doublet of light degenerate quarks (with masses to be identified with the mean up and down quark mass) and a third, single flavour representing the strange quark. The masses employed in our numerical calculations differ from their physical counterparts in nature, which requires approaching the physical point through a chiral extrapolation to the physical quark mass values corresponding to QCD with $2+1$ flavours.

In practice, the tuning of the parameters of the theory is guided by the computation of meson masses, where the physical point is defined by the pion and kaon (as well as some charmed meson) masses taking their physical values.
As the sum of sea quark masses is held constant in our setup, the strange quark mass reaches its physical value from below. However, we would like to anticipate already now that no significant dependence on the light quark masses is observed in our analysis of observables in the charm sector under study.

The charm quark is included as partially quenched valence quark in our calculation, which means that it is not incorporated as a dynamical degree of freedom in the lattice action governing the path integral of the theory. The size of the systematic uncertainty induced by this approximation has been examined in Ref.~\cite{Cali:2019enm} and was found to be, owing to the decoupling of heavier quarks, of the order of a few per cent at most. On the other hand, the comparison of lattice QCD charm quark determinations relying on $2+1$ \cite{McNeile:2010ji, Yang:2014sea, Nakayama:2016atf, Maezawa:2016vgv, Petreczky:2019ozv} and $2+1+1$ dynamical flavours \cite{Carrasco:2014cwa, Alexandrou:2014sha, Chakraborty:2014aca, Bazavov:2018omf, Lytle:2018evc, Hatton:2020qhk}, collected and reviewed in Ref.~\cite{Aoki:2019cca} as far as available at that time, demonstrates an excellent agreement. As for the effects of QED and isospin breaking, the statistical error on our final result will turn out to be substantially larger than the level, at which one expects their influence to affect current lattice QCD computations; this justifies to neglect them here.

Lattice QCD offers a sound, first principles definition of quark masses that derives from a lattice version of the PCAC, viz., partially conserved axial current relation, holding in the chiral symmetry realm of QCD.
Once bare quark masses based on the PCAC relation on the lattice are extracted (as we do it here in case of the charm quark mass), they need to be properly renormalised for their continuum limit to exist.
A convenient object to consider is the renormalisation group invariant (RGI) quark mass, as it is a renormalisation scheme- and scale-independent quantity (if the renormalisation conditions are imposed at zero quark mass \cite{Weinberg:1951ss}), and the determination of the RGI charm quark mass is thus the main goal of this paper.
For connecting the bare (PCAC) mass at a specified low-energy scale with the RGI mass in the continuum limit, we can draw on the study exposed in Ref.~\cite{Campos:2018ahf}, where the renormalisation group running of the quark mass over a wide range of scales for the present three-flavour discretisation of QCD was performed with full non–perturbative precision in the Schrödinger functional scheme, following the strategy of Ref.~\cite{Capitani:1998mq}.
The results of Ref.~\cite{Campos:2018ahf} split into a factor accounting for the universal (i.e., regularisation and flavour independent) running of the quark masses and an associated renormalisation factor at a low-energy scale in an appropriate hadronic scheme, defined via the Schrödinger functional. 
Moreover, our computations involve non-perturbatively determined improvement coefficients and renormalisation constants from Refs.~\cite{Bulava:2015bxa, DallaBrida:2018tpn, Campos:2018ahf, deDivitiis:2019xla, Heitger:2021bmg}, implementing Symanzik improvement for removal of discretisation effects from correlation functions and chiral Ward identities for finite normalisations of quark bilinears (such as the axial vector current) in the $\mathrm{O}(a)$ theory.
Therefore, leading discretisation errors of $\mathrm{O}(a^2)$ in the bulk (except for $\mathrm{O}(g_0^2a)$ uncertainties at the time boundaries) are encountered when taking the continuum limit of correlation functions and renormalised quark masses deduced from them.

The combination of non-perturbative $\mathrm{O}(a)$ improvement and the range of lattice spacings from $\approx 0.087\,$fm down to very fine resolutions of $\approx 0.039\,$fm covered in our work (where the calibration of the lattice spacing to its physical value was done through the intermediate hadronic gradient flow scale $t_0$ in Ref.~\cite{Bruno:2016plf}) allow us to carry out a controlled continuum limit extrapolation, together with the chiral extrapolation in a joint global fit along a trajectory of constant physics towards the physical point of the theory.
In this context, the use of different lattice definitions of the charm quark mass, which distinguish themselves considerably in their cutoff effects, proves to be very valuable, because we are able to confirm that no significant systematic uncertainties are introduced by the global extrapolation to the joint chiral and continuum limit. We find that linear cutoff effects are absent (as expected), but those of $\mathrm{O}(a^2)$ and higher give relevant contributions already at lattice spacings around $0.06\,$fm.

The structure of this paper is as follows. In Section \ref{s:quarkmassesonthelattice}, we introduce the three different definitions of the bare quark mass on the lattice employed and explain their renormalisation pattern in the $\mathrm{O}(a)$ improved theory.
Section \ref{s:quarkmasscomputation} specifies the set of gauge field configuration ensembles and how they realise, in lattice parameter space, the approach to the point of physical quark masses (fixed, in practice, by physical values of pion, kaon and suitable charmed meson masses) along a chiral trajectory and in the continuum limit. There, we also detail the computation of two-point correlation functions, as well as the extraction of meson and bare quark masses from them, and provide the expressions from which the renormalised charm quark mass is obtained and eventually translated into the associated RGI mass.
The joint global chiral-continuum extrapolation of the resulting estimates at still unphysical meson masses and finite lattice spacings are then thoroughly described in Section \ref{s:chiralcontinuum}, where we also elucidate the model averaging procedure that is adopted to arrive at our final result and to quantify its systematic uncertainties.
Our final result on the RGI charm quark mass is presented in Section \ref{s:results}, together with a careful discussion of the various contributions from the single steps of our calculation to its full error budget.
In Section \ref{s:discussion}, we conclude by summarising our findings and compare our final value for the charm quark mass with existing ones from other lattice QCD determinations, see also Appendix \ref{a:running}, where the conversion through perturbative five-loop running to other conventional energy scales in the $\overline{\mathrm{MS}}$ renormalisation scheme is outlined, which represents the customary reference scheme in phenomenological applications.
Lastly, results from several intermediate steps of the analysis, such as meson and (bare) quark masses in lattice units, are tabulated in Appendix~\ref{app:result_tables}.

A preliminary account of our work has been given in Ref.~\cite{Heitger:2019ioq}.
The determination of the mass-degenerate light and strange quark masses on $(2+1)$-flavour CLS ensembles, along the same renormalisation strategy and partly similar analysis methods, was presented in Ref.~\cite{Bruno:2019xed}.

%% file: 02_theory.tex

\section{Quark masses on the lattice with Wilson fermions} \label{s:quarkmassesonthelattice}
Our determination of the physical charm quark mass is based on various different definitions of the quark mass in the formulation of lattice QCD with Wilson quarks. We begin our discussion of these definitions with the bare subtracted quark mass of flavour $i$, which is defined as
\begin{align}
m_{\mathrm{q}, i} = m_{0,i}-m_\mathrm{cr} \equiv \frac{1}{2a}\left(\frac{1}{\kappa_i}-\frac{1}{\kappa_\mathrm{cr}}\right)\,, \label{e:bare_subtracted_quark_mass}
\end{align}
where $\kappa_i$ is the corresponding hopping parameter and $\kappa_\mathrm{cr}$ is the critical hopping parameter defined from vanishing current quark masses in the $SU(N_\mathrm{f})$ symmetric limit. The additive quark mass renormalisation via $m_\mathrm{cr}$ is required due to the loss of chiral symmetry in the Wilson fermion action. The average of two subtracted quark masses of flavours $i$ and $j$ is
\begin{align}
	m_{\mathrm{q}, ij} \equiv \frac{1}{2} (m_{\mathrm{q}, i} + m_{\mathrm{q}, j})\,.
\end{align}

Since composite operators built from Wilson fermion fields show considerable cutoff effects that are enhanced at the scale of the charm quark, $\mathrm{O}(a)$ improvement \`{a} la Symanzik drastically improves their approach towards the continuum limit. A detailed discussion of the underlying improvement pattern in the case of three flavours of quarks can be found in Ref.~\cite{Bhattacharya:2005rb}. We repeat the expression for the renormalised and improved subtracted quark mass,
\begin{align}
m_{i, \mathrm{R}} \equiv\;& Z_\mathrm{m}\left\{\left[m_{\mathrm{q},i} + (r_\mathrm{m}-1)\frac{\mathrm{Tr}\left[M_\mathrm{q}\right]}{N_\mathrm{f}}\right] + a B_i\right\}\,,\label{e:renormalised_subtracted_quark_mass}\\
B_i =\;& b_\mathrm{m}m^2_{\mathrm{q},i}
+ \bar{b}_\mathrm{m}m_{\mathrm{q},i}\mathrm{Tr}\left[M_\mathrm{q}\right] \nonumber\\
\,&+ (r_\mathrm{m}d_\mathrm{m}-b_\mathrm{m})\frac{\mathrm{Tr}\big[M_\mathrm{q}^2\big]}{N_\mathrm{f}}\\
\,&+ (r_\mathrm{m}\bar{d}_\mathrm{m}-\bar{b}_\mathrm{m})\frac{\big(\mathrm{Tr}\left[M_\mathrm{q}\right]\big)^2}{N_\mathrm{f}}\,, \nonumber
\end{align}
with the sea quark mass matrix
\begin{align}
M_\mathrm{q} = \mathrm{diag}(m_{\mathrm{q}, 1}, m_{\mathrm{q}, 2}, \dots, m_{\mathrm{q}, N_\mathrm{f}})\,,
\end{align}
and the mass renormalisation constant $Z_\mathrm{m}$ that not only depends on the bare coupling, but also explicitly on the renormalisation scheme and associated renormalisation scale $a\mu$.\footnote{For ease of notation, we refrain from explicitly writing out corrections of $\mathrm{O}(a^2)$ or higher.}We note in passing that, for achieving $\mathrm{O}(a)$ improvement in a
mass independent scheme, the bare coupling $g_0^2$ has to be improved with
a mass dependent term, $\tilde g_0^2 \equiv g_0^2 (1+b_\mathrm{g}\frac{1}{N_\mathrm{f}}a\mathrm{Tr}\left[M_\mathrm{q}\right])$ as introduced in Ref.~\cite{Luscher:1996sc}.
Yet the chiral trajectory in the quark mass plane associated with
our ensembles (see Section \ref{s:quarkmasscomputation} below) is constructed such as to imply a constant $\tilde g_0^2$ at fixed inverse coupling $\beta=6/g_0^2$ \cite{Bruno:2014jqa}, up to negligible violations \cite{Bruno:2016plf}, irrespective of the knowledge of the improvement coefficient $b_\mathrm{g}$. Therefore, it is legitimate to utilise renormalisation constants as functions of $g_0^2$. 
As apparent from eq.~(\ref{e:renormalised_subtracted_quark_mass}), the subtracted quark mass receives a shift from non-zero sea quark masses which scales with $(r_\mathrm{m}-1)$, where $r_\mathrm{m}$ is the ratio of flavour singlet and non-singlet scalar density renormalisation constants. The cutoff effects at order $\mathrm{O}(a)$ are mass dependent and get fully cleared if the coefficients of the corresponding terms are fixed non-perturbatively.

As an alternative definition of the quark mass on the lattice we employ the bare current quark mass entering the PCAC (partially conserved axial current) relation. For two distinct flavours $i\neq j$ it reads 
\begin{align}
m_{ij} = \frac{\langle (\tilde{\partial}_0 A_0^{ij}(x) + ac_\mathrm{A} \partial_0^* \partial_0^{} P^{ij}(x))P^{ji}(y)\rangle }{2 \langle P^{ij}(x)P^{ji}(y)\rangle}\,, \label{e:m_pcac}
\end{align}
in terms of the improved axial vector current
\begin{align}
(A_\mathrm{I})_\mu^{ij} &= A_\mu^{ij} + ac_\mathrm{A}\partial_\mu P^{ij}\,,
\end{align}
and the pseudoscalar density $P^{ij}$, which in the presence of non-degenerate masses are off-diagonal bilinear fields.
The standard choice for the discretised derivatives is given by the forward and backward differences
\begin{align}
	a\partial_\mu f(x) &\equiv f(x+a\hat\mu) - f(x)\,,& a\partial_\mu^\ast f(x) &\equiv f(x) - f(x-a\hat\mu)\,, \label{e:standard_derivatives}
\end{align}
while $\tilde{\partial}$ denotes the average of these two. Following Refs.~\cite{deDivitiis:1997ka,Guagnelli:2000jw, deDivitiis:2019xla}, we also adopt another realisation of discretised derivatives based on the substitutions
\begin{align}
	\tilde{\partial}_\mu &\rightarrow \tilde{\partial}_\mu \left(1-\tfrac{1}{6} a^2 \partial^*_\mu \partial^{}_\mu\right)\,, &
	\partial^*_\mu \partial^{}_\mu &\rightarrow \partial^*_\mu \partial^{}_\mu \left(1 - \tfrac{1}{12} a^2 \partial^*_\mu \partial^{}_\mu\right)\,. \label{e:improved_derivatives}
\end{align}
Since the discretisation effects introduced by these derivatives acting on smooth functions are of $\mathrm{O}(a^4)$, we will refer to this second choice as improved derivatives.

Upon renormalisation, eq.~(\ref{e:m_pcac}) leads to the renormalised $\mathrm{O}(a)$ improved current quark mass $m_{\mathrm{R}, ij}$,
\begin{align}
m_{\mathrm{R}, ij}= \frac{Z_\mathrm{A}}{Z_\mathrm{P}} m_{ij} \left[1+(b_\mathrm{A}-b_\mathrm{P}) am_{\mathrm{q},ij} + (\bar{b}_\mathrm{A}-\bar{b}_\mathrm{P})a \mathrm{Tr}\left[M_\mathrm{q}\right]\right]\,, \label{e:mpcac_renormalised}
\end{align}
which in the theory with mass non-degenerate quarks is equivalent to the average of two renormalised current quark masses of flavour $i$ and $j$.
Here, $Z_\mathrm{A}(\tilde{g}_0^2)$ and $Z_\mathrm{P}(\tilde{g}_0^2, a\mu)$ label the renormalisation constants of the axial current and the pseudoscalar density, respectively, the latter carrying the renormalisation scheme and scale dependences of the quark mass as does $Z_\mathrm{m}$ in eq.~(\ref{e:renormalised_subtracted_quark_mass}) above.

We can now equate both expressions for the renormalised quark mass, eqs.~(\ref{e:renormalised_subtracted_quark_mass}) and (\ref{e:mpcac_renormalised}), to derive
\begin{align}
m_{\mathrm{q}, ij} = \frac{m_{ij}}{Z}-(r_\mathrm{m}-1) \frac{\mathrm{Tr}\left[M_\mathrm{q}\right]}{N_\mathrm{f}} + \mathrm{O}(am_{ij},a\mathrm{Tr}\left[M_\mathrm{q}\right])\,, \label{e:mqij_from_mij}
\end{align}
where $Z$ is the scale independent combination of renormalisation constants
\begin{align}
Z (\tilde{g}_0^2) \equiv \frac{Z_\mathrm{m}(\tilde{g}_0^2, a\mu)Z_\mathrm{P}(\tilde{g}_0^2, a\mu)}{Z_\mathrm{A}(\tilde{g}_0^2)}\,, \label{e:Z_definition}
\end{align}
and employ this relation to eliminate the bare subtracted quark mass in eq.~(\ref{e:mpcac_renormalised}) in favour of PCAC masses to obtain
\begin{align}
m_{\mathrm{R}, ij}=\;& \frac{Z_\mathrm{A}}{Z_\mathrm{P}} m_{ij} \Bigg[1+(\tilde{b}_\mathrm{A}-\tilde{b}_\mathrm{P}) am_{ij}  \vphantom{\frac{A}{B}} \label{e:mpcac_renormalised_rm2} \\
& + \left( \frac{(\bar{b}_\mathrm{A}-\bar{b}_\mathrm{P})}{Zr_\mathrm{m}} - (\tilde{b}_\mathrm{A}-\tilde{b}_\mathrm{P}) \frac{(r_\mathrm{m}-1)}{r_\mathrm{m}N_\mathrm{f}} \right) a M_\mathrm{sum}\Bigg]\,, \nonumber
\end{align}
with
\begin{align}
\tilde{b}_\mathrm{A}-\tilde{b}_\mathrm{P} &\equiv \frac{(b_\mathrm{A}-b_\mathrm{P})}{Z}\,, \\
M_\mathrm{sum} &\equiv m_{12} + m_{23} + \dots + m_{(N_\mathrm{f}-1)N_\mathrm{f}} + m_{N_\mathrm{f}1} \nonumber\\
&= Zr_\mathrm{m}\mathrm{Tr}\left[M_\mathrm{q}\right] + \mathrm{O}(a\mathrm{Tr}\left[M_\mathrm{q}\right])\,.
\end{align}
Note that in eq.~(\ref{e:mpcac_renormalised_rm2}) all dependence on $m_\mathrm{cr}$ has been removed.

Yet another definition of the renormalised quark mass with different cutoff effects may be constructed along the so-called ratio-difference method \cite{Durr:2010aw}, where we again combine the two definitions for renormalised quark masses, eqs.~(\ref{e:renormalised_subtracted_quark_mass}) and~(\ref{e:mpcac_renormalised}), but in a different way than before. Namely, we introduce a ratio of current quark masses, $r$, and a difference of subtracted quark masses, $d$, through
\begin{align}
r_{ij} &\equiv \frac{m_{ii'}}{m_{jj'}}\,, & d_{ij} &\equiv am_{\mathrm{q}, i} - am_{\mathrm{q}, j}\,,
\end{align}
where $i$ and $i'$ label two distinct but mass degenerate quark flavours. This setup is chosen such that the multiplicative renormalisation of the current quark masses cancels in the ratio, while the additive renormalisation of the bare subtracted quark masses (via $m_\mathrm{cr}$) and the leading-order dependence on $\mathrm{Tr}\left[M_\mathrm{q}\right]$ drop out in the difference. As explained in detail in Ref.~\cite{Heitger:2020mkp}, the $\mathrm{O}(a)$ improved versions of $r_{ij}$ and $d_{ij}$ can be written as
\begin{align}
r_{\mathrm{I},ij} &= r_{ij} \left[ 1+ (b_\mathrm{A}-b_\mathrm{P}) d_{ij}\right]\,, \label{e:bmw_improved_ratio}\\
d_{\mathrm{I},ij} &= d_{ij} \left[1 + b_\mathrm{m} d_{ij} \frac{r_{ij}+1}{r_{ij}-1} + b_\mathrm{m}d_{ij} \frac{2(1-r_\mathrm{m})M_\mathrm{sum}}{r_\mathrm{m}m_{jj'}(r_{ij}-1)N_\mathrm{f}} + a \bar{b}_\mathrm{m}\frac{M_\mathrm{sum}}{Zr_\mathrm{m}}\right]\,. \label{e:bmw_improved_difference}
\end{align}
From the latter, an estimator for the renormalised quark mass is formed by
\begin{align}
m_{i, \mathrm{R}}^{(\mathrm{rd})} = Z_\mathrm{m} \frac{r_{\mathrm{I},ij}\,d_{\mathrm{I},ij}}{r_{\mathrm{I},ij}-1}\,, \label{e:ratio_difference}
\end{align}
with the mass renormalisation factor $Z_\mathrm{m}$ already introduced in eq.~(\ref{e:renormalised_subtracted_quark_mass}).

For the improvement and renormalisation of the axial current we use the improvement coefficient $c_\mathrm{A}$ calculated in Ref.~\cite{Bulava:2015bxa} and the values of $Z_\mathrm{A}(g_0^2)$ determined within the chirally rotated Schrödinger functional \cite{DallaBrida:2018tpn}. The scale dependent renormalisation constant $Z_\mathrm{P}$ was computed in Ref.~\cite{Campos:2018ahf}, together with the universal factor that relates quark masses at finite renormalisation scale to their RGI counterparts. The renormalisation constant $Z$ defined in eq.~(\ref{e:Z_definition}) was calculated for our setup in Ref.~\cite{deDivitiis:2019xla}. It can also be extracted exploiting the numerical results on $Z_\mathrm{S}/Z_\mathrm{P}$ presented in Ref.~\cite{Heitger:2020mkp}.

The non-perturbative determination of the improvement coefficients $(b_\mathrm{A}-b_\mathrm{P})$ and $b_\mathrm{m}$ has been performed in Ref.~\cite{deDivitiis:2019xla}, along with $Z$, for two different lines of constant physics (LCP) realised in the Schrödinger functional framework for the range of lattice spacings relevant here. More precisely, there is a set of coefficients computed on an LCP adapted for use in a fully massless setup (called `LCP-0' in \cite{deDivitiis:2019xla}), whereas, by contrast, a further set of estimates (labelled `LCP-1') for $Z$, $(b_\mathrm{A}-b_\mathrm{P})$ and $b_\mathrm{m}$ refers to an LCP at fixed heavy quark mass in the valence sector. Therefore, particularly the latter appears to be predestined for applications with heavy (valence) quarks, because mass dependent cutoff effects of all orders are expected to be accounted for in the improvement coefficients and renormalisation constant from this LCP. In fact, the investigations in Refs.~\cite{Heitger:2003ue, Fritzsch:2010aw} have shown that the scaling of meson observables involving heavy quarks towards the continuum limit benefits from including these mass dependent cutoff effects in the LCP definition of the improvement coefficients and renormalisation factor in question. This conjecture will be tested later, when we discuss the chiral-continuum extrapolation of our results. 

Non-perturbative results for $r_\mathrm{m}$ for the full range of couplings recently became available from the determination in Ref.~\cite{Heitger:2021bmg} (earlier findings for a subset of the CLS ensembles are available in Ref.~\cite{Bali:2016umi}). We employ these for the corresponding improvement terms in eqs.~(\ref{e:mpcac_renormalised_rm2}) and (\ref{e:bmw_improved_difference}) and observe only a marginal influence on heavy quark masses at finite lattice spacing.

Finally, for the improvement coefficient $(\bar{b}_\mathrm{A}-\bar{b}_\mathrm{P})$ and $\bar{b}_\mathrm{m}$ multiplying cutoff effects depending on the sea quark masses, non-perturbative results with errors of $\mathrm{O}(100\%)$ were obtained in Refs.~\cite{Korcyl:2016cmx, Korcyl:2016ugy} using coordinate-space methods. Given the large uncertainties in these determinations and the fact that the coefficients are of $\mathrm{O}(g_0^4)$ in perturbation theory (as opposed to $\mathrm{O}(g_0^2)$ for the coefficients $b_X$ mentioned before), we ignore the corresponding terms of $\mathrm{O}(a\mathrm{Tr}\left[M_\mathrm{q}\right])$ in our improvement procedure. We note that the associated contributions from the sea quark masses are highly suppressed compared to the heavy valence quark mass effects that are removed non-perturbatively.

%% file: 03_computations.tex
\section{Computational details on the quark mass extraction}\label{s:quarkmasscomputation}
The numerical study described in the following has been carried out on the $(2+1)$-flavour CLS gauge field configuration ensembles described in Refs.~\cite{Bruno:2014jqa,Mohler:2017wnb}. Gluons are implemented via the tree-level improved Lüscher-Weisz gauge action \cite{Luscher:1984xn} and quarks are treated as clover improved Wilson fermions \cite{Sheikholeslami:1985ij}. All ensembles employed in our computations feature open boundary conditions in time direction to prevent a freezing of the topological charge \cite{Luscher:2011kk, Luscher:2012av}. Twisted mass reweighting \cite{Luscher:2008tw} has been applied in the simulations for the doublet of light sea quarks and the RHMC algorithm \cite{Kennedy:1998cu,Clark:2006fx} has been used for the single strange quark in the sea. 

Charm quarks are introduced in the valence sector only, rendering our theory a partially quenched approximation of QCD. Whereas the effect of a dynamical charm quark in low-energy purely gluonic (and light flavour) observables is expected to be suppressed below the available accuracy \cite{Bruno:2014ufa}, this may be different for observables at the scale of the charm quark, and therefore a systematic uncertainty is present owing to neglecting a dynamical charm in the sea in our final result for the physical charm quark mass. A conservative upper bound of $5\%$ for the relative size of this uncertainty was estimated in Refs.~\cite{Cali:2018owe, Cali:2019enm} from calculations with two dynamical charm quarks. 

Ensembles with five different lattice spacings in a range from $0.087\,$fm down to $0.039\,$fm, together with non-perturbative $\mathrm{O}(a)$ improvement, allow for a controlled continuum limit extrapolation. Moreover, these ensembles cover pion masses in the range $\left[200\,\mathrm{MeV},\, 420\,\mathrm{MeV}\right]$ such that the dependence of our results for the charm quark mass on the light quark masses may be carefully investigated. In general, however, the effect of unphysically large sea quark masses on observables at the scale of the charm quark is expected to be small. We perform our calculations on a set of ensembles, for which the sum of the bare sea quark masses $\mathrm{Tr}\left[M_\mathrm{q}\right]$ is held fixed. In such a setup, the strange quark mass approaches its physical value from below when the light quark masses are lowered towards their physical values. 

The strategy to fix an LCP trajectory in bare lattice parameter space within the $\mathrm{Tr}\left[M_\mathrm{q}\right]=\mathrm{const.}$ CLS ensembles and its exact specification are explained in great detail in Refs.~\cite{Bruno:2016plf, Bruno:2019vup}. Since a fixed value of the bare trace of the quark mass matrix does not imply a constant renormalized trace of the quark mass matrix, the generated gauge field configuration ensembles do not lie strictly on an LCP. Therefore, and since the deviations of $\mathrm{Tr}\left[M_\mathrm{R, q}\right]$ from a constant value have been observed to be larger than what would be expected from $\mathrm{O}(a)$ cutoff effects \cite{Bruno:2016plf}, the chiral trajectory has been redefined in terms of a constant value of 
\begin{align}
	\phi_4 \equiv 8 t_0 \left(m_\mathrm{K}^2 + \frac{1}{2}m_\pi^2\right)\,, \label{e:phi_4}
\end{align}
where the gluonic quantity $t_0$ is defined from the Wilson flow \cite{Luscher:2010iy} and $m_\pi$ and $m_\mathrm{K}$ denote the masses of the lightest mesons composed of light and light-strange quarks, respectively.
To approach the physical point in a chiral extrapolation, the value of $\phi_4$ on each ensemble has to be fixed to the physical value
\begin{align}
	\phi_4^\mathrm{phys} =  8 t_0^\mathrm{phys} \left[\left(m_\mathrm{K}^\mathrm{phys}\right)^2 + \frac{1}{2}\left(m_\pi^\mathrm{phys}\right)^2\right] = 1.120(24)\,,
\end{align}
based on the physical values of pion and kaon masses in isospin symmetric QCD as detailed in Ref.~\cite{Aoki:2016frl} and the physical value of $t_0$ determined in Ref.~\cite{Bruno:2016plf}. In this appropriately adapted setup, the dependence of physical observables on the sea quark masses can now be parametrised via 
\begin{align}
	\phi_2 \equiv 8 t_0 m_\pi^2 \label{e:phi_2}\,,
\end{align}
while the physical point is recovered by an extrapolation to 
\begin{align}
	\phi_2^\mathrm{phys} = 8 t_0^\mathrm{phys} \left(m_\pi^\mathrm{phys}\right)^2 = 0.0804(18)\,.
\end{align}

To ensure $\phi_4 = \phi_4^\mathrm{phys}$ on each ensemble, we resort to small quark mass shifts onto the chiral trajectory by means of a Taylor expansion of the entering correlation functions as described in Ref.~\cite{Bruno:2016plf}. Following this reference, the derivative of an observable $\mathcal{O}$ with respect to a change in the quark mass $m_{\mathrm{q}, i}$ of flavour $i$ is given by
\begin{align}
\frac{\mathrm{d} \langle\mathcal{O}\rangle}{\mathrm{d} m_{\mathrm{q}, i}} =
\left\langle \frac{\partial \mathcal{O}}{\partial m_{\mathrm{q}, i}} \right\rangle
- \left\langle \mathcal{O} \frac{\partial S}{\partial m_{\mathrm{q}, i}} \right\rangle
+  \left\langle\mathcal{O}\right\rangle \left\langle \frac{\partial S}{\partial m_{\mathrm{q}, i}} \right\rangle\,. \label{e:mass_shifts}
\end{align}
In order to carry out this shift at the level of the correlation functions employed in our work, their partial derivatives were numerically evaluated (simultaneously with the calculation of the correlators themselves) utilising the \texttt{mesons} program package \cite{mesons}, augmented by 
sets of measurements for the derivative terms of the action with respect to the quark mass in (\ref{e:mass_shifts}) that we had to extend compared to what was already available as a result of \cite{Bruno:2016plf}. The shift of observables depending on $N_\mathrm{f}=2+1$ quark masses is then accomplished by the replacement
\begin{align}
\langle \mathcal{O} \rangle \rightarrow \langle \mathcal{O}\rangle +\sum_{i=1}^{N_\mathrm{f}}\Delta m_{\mathrm{q},i}\frac{\mathrm{d} \langle\mathcal{O}\rangle}{\mathrm{d} m_{\mathrm{q}, i}}\,, \label{e:shifted_observable}
\end{align}
and we choose the same value of $\Delta m_{\mathrm{q},i}$ for all three quark flavours. In practice, the value of $\Delta m_{\mathrm{q},i}$ on each ensemble was obtained in an iterative procedure that stops when the shifted $\phi_4$ matches its physical value well below the statistical precision.

\subsection{Correlation functions and fixing of the physical charm quark mass}\label{ss:cfs+fixingphysmcharm}
As already indicated above, we have performed correlator measurements on 15 ensembles on the $\mathrm{Tr}\left[M_\mathrm{q}\right]=\mathrm{const.}$ trajectory from the $(2+1)$-flavour CLS gauge configuration data base, using the \texttt{mesons} code~\cite{mesons}. The specifications of these ensembles and their associated statistics in the number of molecular dynamic units (MDU) are summarised in  Table \ref{t:measurements}.
To extract ground state meson masses as well as quark masses through the PCAC relation, we consider zero-momentum two-point correlation functions defined by
\begin{align}
f_{OO'}^{rs}(x_0,y_0)=-\frac{a^6}{L^3}\sum_{\mathbf{x},\mathbf{y}}\left\langle O^{rs}(x_0,\mathbf{x})O^{\prime, sr}(y_0,\mathbf{y})\right\rangle\,, \label{e:mcharm_fOO}
\end{align}
where $r$ and $s$ are flavour indices. $y_0$ denotes the time coordinate of the source, i.e., the time slice where the source has a non-vanishing norm, and $x_0$ is the time coordinate of the sink. The operators $O^{rs}$ are quark bilinear covariant fields composed as
\begin{align}
O^{rs}(x)=\bar{\psi}^r(x)\,\Gamma\,\psi^s(x)\,, \label{e:mcharm_Ors}
\end{align}
in terms of $\Gamma$, representing the combination of Euclidean gamma matrices that read $\gamma_0\gamma_5$ for the time component of the axial current, $A_0$, and $\gamma_5$ for the pseudoscalar density, $P$.\footnote{While also $\gamma_k$ for the spatial components of the vector current $V_k$ was included in the set of evaluated correlators, the vector meson mass does not enter our final analysis, see below.} 
Since translational invariance in time direction is broken by the open boundary conditions, we fix the temporal source positions $y_0$ to the two time slices at $a$ and $T-a$ as proposed in Ref.~\cite{Bruno:2014lra}. To achieve good statistical accuracy, we average over 16 $\mathrm{U}(1)$ noise sources per source position and exploit time reversal symmetry.

The correlation functions $f_{OO'}^{rs}$ were evaluated for all combinations of light and strange as well as for two choices of heavy valence quark flavours. For the latter, the hopping parameters have been chosen such that they encompass the supposed hopping parameter of the physical charm quark $\kappa_\mathrm{c}$ sufficiently close. An overview of these hopping parameters is given in Table \ref{t:kappa_h}. Distance preconditioning \cite{deDivitiis:2010ya}, in its variant as sketched in Ref.~\cite{Collins:2017iud} (and implemented in the \texttt{mesons} measurement code~\cite{mesons}), has been employed for the calculation of the heavy quark propagators, in order to ensure the stability and accuracy of the associated numerical inversions of the Dirac operator across the complete temporal extent of the lattice. 

\begin{table}[t]
	\setlength{\abovecaptionskip}{10pt}
	\setlength{\belowcaptionskip}{0pt}
	\centering\small
	\renewcommand{\arraystretch}{1.25}
	\caption{Overview of the gauge field configuration ensembles (labelled by `id') used in this study. We list the geometry ($T\times L^3$) and the bare parameters of the action as well as the resulting approximate lattice spacings and meson masses. The statistics included is quantified by the number of molecular dynamics units $N_\mathrm{MDU}$. We performed measurements every $4\,$MDU except for J303 and J500, where the spacing between two measurements is $8\,$MDU. The exponential autocorrelation time is defined from $t_0$ as given by eq.~(\ref{e:tau_exp}).}
	\input{./tables/tab_ensembles.tex}
	\label{t:measurements}
\end{table}

We follow two different strategies to tune to the physical charm quark mass. Firstly, since $\mathrm{Tr}\left[M_\mathrm{q}\right]$ is held constant in our framework, the flavour averaged meson mass
\begin{align}
m_{\bar{\mathrm{D}}} \equiv \frac{2}{3} m_\mathrm{D} + \frac{1}{3} m_{\mathrm{D}_\mathrm{s}} \,,
\end{align}
built from $D$ and $D_\mathrm{s}$ mesons, is approximately constant as well. At the same time, the statistical precision of the corresponding effective masses is convincingly good. Secondly, we consider the effective mass of the connected part of the $\eta_\mathrm{c}$ meson, called $m_{\eta_\mathrm{c}}$ from now on. The signal of this observable is clean, and the effect of neglecting the disconnected part is expected to be subleading \cite{deForcrand:2004ia}, particularly so for singlet mesons composed of heavy quarks.
Hence we have performed all measurements including heavy quarks for both charmed hopping parameters and, to eventually obtain the physical charm quark mass, interpolate the meson mass results to their respective physical, i.e., experimental values jointly with a chiral-continuum extrapolation procedure, as will be explained in detail in the next section. Let us mention that we also had tested the spin-flavour averaged meson mass constructed from heavy-light and heavy-strange pseudoscalar and vector mesons \cite{Heitger:2019ioq} to fix the charm quark mass.\footnote{Since the spin-dependent first-order correction of HQET to the meson mass is cancelled in this average \cite{Falk:1992wt, Neubert:1993mb}, one might a priori expect reduced cutoff effects from the use of this average.} However, even though one is able to extract the ground states of the vector mesons (despite the correlators being considerably noisier than in the pseudoscalar channel) on all ensembles, we did not see any improvement in the approach of our results to the continuum limit compared to the other two methods. Moreover, due to the still relatively large statistical uncertainties of the vector meson masses, there is also no gain in overall statistical precision such that we did not include the spin-flavour averaged meson mass in our final analysis.

\begin{table}
	\setlength{\abovecaptionskip}{10pt}
	\setlength{\belowcaptionskip}{0pt}
	\centering\small
	\renewcommand{\arraystretch}{1.25}
	\caption{Hopping parameters of the partially quenched heavy valence quarks in the charm region serving to implicitly fix $\kappa_\mathrm{c}$, provided by \protect\cite{heavy_kappas} in the context of determining leptonic decay constants of $D$ and $D_\mathrm{s}$ mesons~\protect\cite{Collins:2017iud, Collins:2017rhi}.
	}
	\input{./tables/tab_kappa_h.tex}
	\label{t:kappa_h}
\end{table}

We carry out our statistical error analysis using the $\Gamma$-method \cite{Wolff:2003sm} deriving from autocorrelation functions, supplemented by automatic differentiation for the error propagation \cite{Ramos:2018vgu}, and take into account the effect of the exponential tails in the autocorrelation functions owing to slow modes \cite{Schaefer:2010hu}. As estimate for the required exponential autocorrelation time on the CLS ensembles we recourse to the formula
\begin{align}
	\tau_\mathrm{exp} = 14(3) \frac{t_0}{a^2} \label{e:tau_exp}
\end{align}
quoted in Ref.~\cite{Bruno:2014jqa} for our set of ensembles. 

\subsection{Extraction of meson and bare PCAC quark masses}
\begin{figure}[hbt!]
	\centering
	\includegraphics[width=\linewidth]{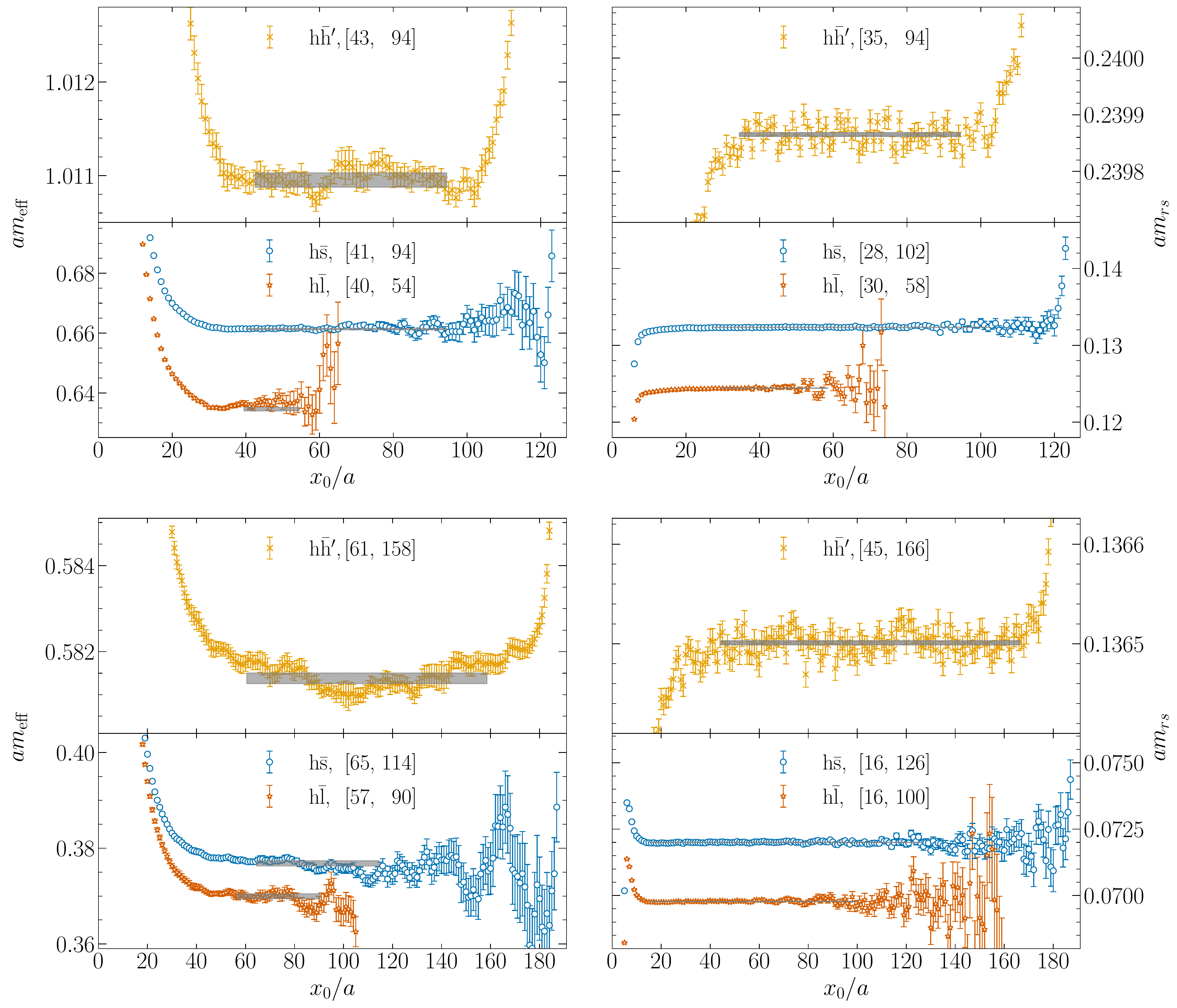}
	\caption{Results from the extraction of meson and quark masses on the most chiral ensemble, D200 (\textit{top}), and one of the ensembles at smallest lattice spacing, J501 (\textit{bottom}). The labels denote the quark content, where h corresponds to $\kappa_{\mathrm{h}_2}$ of Table \ref{t:kappa_h}, and the intervals $[t_\mathrm{min}/a,t_\mathrm{max}/a]$ indicate where the masses develop plateaux. \textit{Left}: Effective meson masses that are employed in the tuning to the physical charm quark mass. The shaded areas indicate the ground state contributions to the two-state fits. \textit{Right}: Bare current quark masses, employing improved derivatives in their lattice definition. The shaded regions indicate the plateau averages, which are too precise to be resolved in a few cases. Excessively fluctuating points based on heavy-light correlation functions are not displayed.}
	\label{fig:masses}
\end{figure}

Ground state masses of pseudoscalar mesons are extracted from the effective mass
\begin{align}
	am_\mathrm{eff}(x_0) = \ln\left(\frac{f_\mathrm{PP}(x_0)}{f_\mathrm{PP}(x_0+a)}\right)\,,
\end{align}
which forms a plateau for sufficiently large source-sink separations. The computation of pion and kaon masses on lattices with open boundaries has been discussed in Refs.~\cite{Bruno:2014jqa, Bruno:2014lra, Bruno:2016plf}, and we employ the same methods to determine the plateau region, i.e., fits to the exponential corrections originating from boundary effects and short source-sink separations. Including the dominant contributions amounts to model the effective masses with an ansatz
\begin{align}
	am_\mathrm{eff}(x_0) = am_\mathrm{PS} \left(1+c_1 \mathrm{e}^{-E_1x_0} + c_2 \mathrm{e}^{-E_\mathrm{2PS}(T-x_0)} \right)\,, \label{e:meff_fit}
\end{align}
where $m_\mathrm{PS}$ gives the mass of the pseudoscalar meson in question, $E_1=m'-m_\mathrm{PS}$ the difference between ground and first excited state energies, and $E_\mathrm{2PS} \approxeq 2m_\mathrm{PS}$. The energies and the parameters $c_i$ are chosen to be free fit parameters.
The same procedure is applied to the effective mass of the $\eta_\mathrm{c}$ meson. For heavy-light and heavy-strange mesons we restrict ourselves to a region where the statistical fluctuations are small such that effects of the boundary at maximal source-sink separation do not affect the quality of the plateau. We show representative effective masses for these mesons, entering into the interpolations to fix the physical charm quark mass, in the left part of Figure~\ref{fig:masses}; there the exponential corrections for small source-sink separations and close to the boundaries are clearly visible.

To specify the optimal fit window, we vary the minimal source-sink distance included in the fit and monitor its quality through the so-called `corrected' $\chi^2$, cf.~eq.~(\ref{e:corrected_chisq}), of an uncorrelated fit. $\chi^2_\mathrm{corrected}$ is based on the correlations present in the data, as motivated and elaborated in Ref.~\cite{Bruno:2020xxxx}. The beginning of the plateau range is set to the time slice where the contribution of the excited state corrections is four times smaller than the statistical error on the effective mass. The upper end of the fit interval is taken from a fit to the boundary corrections in case of mesons with constant signal to noise ratio, while it is a bound of 3\% on the relative error of the correlation function in case of heavy-light mesons. The meson mass can then be defined either as the weighted plateau average within the plateau range thus determined or directly from the fit to the plateau and its exponential corrections according to eq.~(\ref{e:meff_fit}). Because of their compatibility with each other, in the final analysis we just opted for the latter. Note that thanks to the use of distance preconditioning the data on heavy quark propagators over the full temporal range can be incorporated in the plateau fits. This effect is actually enhanced on the J ensembles close to the continuum limit, which have a maximal source-sink separation of $189a$.

As an identity on the operator level, the PCAC relation is valid on every time slice and, consequently, current quark masses exhibit a plateau, too. Contact terms lead to deviations from this plateau behaviour for small source-sink separations, as do cutoff effects close to the boundary. Following Ref.~\cite{Bruno:2016plf}, we estimate the time slice extent of these deviations with the help of the same methods as for the effective meson masses above, i.e., modeling the corrections by exponential fits. In identifying plateaux, we ensure to extract the mass values in the bulk of the lattice where $\mathrm{O}(g_0^2a)$ cutoff effects from the open boundaries are absent. For heavy-light and heavy-strange quark masses, the signal is lost in noise for large source-sink separations. Typical current quark mass plateaux are displayed in the right part of Figure~\ref{fig:masses}.

In Appendix~\ref{app:result_tables} we tabulate the ensemble-wise results for the meson masses and the different bare current quark masses. In each table we list results before and after the mass shifting procedure along the prescription in eq.~(\ref{e:shifted_observable}).
Accounting for these mass shifts increases the statistical uncertainties of the meson and bare quark masses considerably. This is due to the fact that the target for the mass shift, $\phi_4^\mathrm{phys}$, is only known with about $1\,$\% precision, dominated by the uncertainty of the physical lattice scale. It should also be noted that the target value of $\phi_4^\mathrm{phys}$ is the same for all ensembles. For this reason the shifted masses from different gauge field ensembles cannot be considered uncorrelated.

\subsection{Quark mass combinations}
The PCAC definition (\ref{e:m_pcac}) amounts to analysing the correlation functions $f^{rs}_{\mathrm{A}_0\mathrm{P}}$ and $f^{rs}_{\mathrm{PP}}$ as described in the foregoing subsection, in order to arrive at numerical estimates for the current quark masses referring to the various possible mass degenerate and non-degenerate combinations of light, strange and heavy quarks. Assuming that a proper interpolation of these intermediate results involving the two different hopping parameter choices in the charm region as a function of one of the two charmed meson masses (viz., $m_{\bar{\mathrm{D}}}$ and $m_{\eta_\mathrm{c}}$, advocated in Subsection~\ref{ss:cfs+fixingphysmcharm} to fix the physical charm quark mass by their respective experimental values) has been performed, one may define the renormalised charm quark mass from \textit{heavy-heavy} correlation functions as
\begin{align}
m_\mathrm{R, c}^{(\mathrm{c})} &= m_\mathrm{R, cc'}\,, \label{e:charm_m_cc}
\end{align}
where $\mathrm{c}$ and $\mathrm{c'}$ denote two distinct but mass degenerate quark flavours and the appropriate renormalisation and improvement pattern is given in eq.~(\ref{e:mpcac_renormalised_rm2}). This definition always provides clean signals and therefore small statistical errors of the plateau averages. The dependence of $m_\mathrm{R, c}^{(\mathrm{c})}$ on the light quark masses (encoded in its pion mass dependence) is small. Since the mass dependent cutoff effects are at the scale of the charm quark mass, we expect them to be sizeable, although effects of $\mathrm{O}(a)$ are cancelled non-perturbatively in our lattice formulation of the theory. 

The size of these cutoff effects is reduced when \textit{heavy-light} and \textit{heavy-strange} correlation functions are employed from which estimators for the mass of the charm quark may then be obtained via
\begin{align}
m_\mathrm{R, c}^{(\mathrm{l})} &= 2 m_\mathrm{R, lc} -m_\mathrm{R, ll'}\,, &
m_\mathrm{R, c}^{(\mathrm{s})} &= 2 m_\mathrm{R, sc} -m_\mathrm{R, ss'}\,. \label{e:heavy_light_pcac_definitions}
\end{align}
As evident in our data, the definitions in eq.~(\ref{e:heavy_light_pcac_definitions}) have a relevant dependence on the sea quark masses which, however, turns out to be significantly reduced in the associated flavour-averaged charm quark mass
\begin{align}
\overline{m}_\mathrm{R, c} = \frac{2}{3} m_\mathrm{R, c}^{(\mathrm{l})} + \frac{1}{3} m_\mathrm{R, c}^{(\mathrm{s})}\,, \label{e:charm_m_lc}
\end{align}
because of the average sea quark mass being held constant along our trajectory in lattice parameter space towards the physical point.
The signal of heavy-light and heavy-strange quark masses deteriorates for large source-sink separations. Accordingly, the statistical error on the charm quark mass definition $\overline{m}_\mathrm{R, c}$ is generally larger than on $m_\mathrm{R, c}^{(\mathrm{c})}$.

As a third option to arrive at a sensible prescription to compute the charm quark's mass, we adopt the \textit{ratio-difference method}, eq.~(\ref{e:ratio_difference}). Here, the first flavour is chosen to be the heavy valence quark and the second flavour is one of the sea quarks. Adhering to the same argument in favour of the linear combination in (\ref{e:charm_m_lc}), we deduce the renormalised charm quark mass from the corresponding flavour average also in this case, i.e.,
\begin{align}
m_\mathrm{R, c}^{(\mathrm{rd})} = Z_\mathrm{m}\left(
\frac{2}{3}\frac{r_{\mathrm{I,cl}}\,d_{\mathrm{I,cl}}}{r_{\mathrm{I,cl}}-1} +
\frac{1}{3}\frac{r_{\mathrm{I,cs}}\,d_{\mathrm{I,cs}}}{r_{\mathrm{I,cs}}-1}\right)\,, \label{e:charm_m_rd}
\end{align}
with the improved ratio $r_{\mathrm{I},ij}$ and difference $d_{\mathrm{I},ij}$ already given in eqs.~(\ref{e:bmw_improved_ratio}) and (\ref{e:bmw_improved_difference}), respectively. The anticipated advantage of this estimator is that systematic uncertainties owing to an imprecise determination of the chiral point, which are naturally associated with subtracted quark masses, are cancelled in the quark mass from the ratio-difference method.

\subsection{Renormalised quark masses} \label{ss:MRGI}
As a central result of this paper, we wish to quote the renormalisation group invariant (RGI) value of the charm quark mass. For a given quark flavour, the RGI quark mass, $M$, is defined through the formally exact expression
\begin{align}
M \equiv m_{\mathrm{R}}(\mu) \left[2b_0 g_\mathrm{R}^2(\mu)\right]^{-\frac{d_0}{2b_0}} \exp \left\{-\int_{0}^{g_\mathrm{R}^2(\mu)}\mathrm{d}g \left[\frac{\tau(g)}{\beta(g)} - \frac{d_0}{b_0g}\right] \right\}\,, \label{e:M_RGI}
\end{align}
where it is assumed that renormalisation conditions are imposed at zero quark mass, which suggests calling those schemes mass independent. This implies ratios of renormalised quark masses for different flavours to become scale and scheme independent constants. In the same way as the QCD $\Lambda$-parameter associated with the running of the renormalised coupling $g_\mathrm{R}(\mu)$, $M$ belongs to the RGI quantities whose total dependence on the renormalisation scale $\mu$ vanishes. In addition we recall that, even though perturbative expansions for the RG group functions $\beta(g)$ and $\tau(g)$ exist, the $\Lambda$-parameter and the RGI quark masses are generically defined independent of perturbation theory and, particularly, the RGI quark masses $M$ are renormalisation scheme independent. Therefore, we consider them ideal for comparisons with results from either experiments or other theoretical determinations. Conventions regarding the RG $\beta$- and $\tau$-functions as well as their respective perturbative coefficients $b_i$ and $d_i$ can be found, e.g., in Refs.~\cite{Capitani:1998mq,DellaMorte:2005kg}.

The renormalised coupling $g_\mathrm{R}(\mu)$ and the continuum RG functions of the running coupling and quark mass entering eq.~(\ref{e:M_RGI}) were accurately obtained for the three-flavour theory at hand by applying the non-perturbative step scaling approach within the Schr\"odinger functional scheme~\cite{Brida:2016flw,DallaBrida:2016kgh,Campos:2018ahf}. Hence, we have all the ingredients available to merge the above definition of $M$ with one of the estimators for the renormalised charm quark mass introduced in the previous subsection and calculate the RGI charm quark mass from it. Explicitly, on the basis of the expressions in eqs.~(\ref{e:charm_m_cc}), (\ref{e:charm_m_lc}) and (\ref{e:charm_m_rd}), their relation to the RGI mass then reads:
\begin{align}
	M_\mathrm{c} = \frac{M}{m_\mathrm{R}(\mu_\mathrm{had})} m_{\mathrm{R}, \mathrm{c}}(\mu_\mathrm{had})\,.
\end{align}
The renormalisation scale dependence of $m_{\mathrm{R}, \mathrm{c}}$ is inherited from the corresponding scale dependence of the respective $Z$-factor (see Section~\ref{s:quarkmassesonthelattice}), which turns the underlying bare quark mass into the renormalised one. Thanks to the determination of the universal, flavour independent ratio 
\begin{align}
	\frac{M}{m_\mathrm{R}(\mu_\mathrm{had})} = 0.9148(88)\,, \label{e:M_over_m_nf3}
\end{align}
in the Schrödinger functional scheme for $N_\mathrm{f}=3$ massless flavours at $\mu_\mathrm{had} = 233(8)\,$MeV~\cite{Campos:2018ahf}, the renormalisation factors and the ratio $M/m_\mathrm{R}$ can be matched at the hadronic scale $\mu=\mu_\mathrm{had}$ such that $M_\mathrm{c}$ is readily returned. Further details concerning the non-perturbative computation of the running factor (\ref{e:M_over_m_nf3}) can be found in Ref.~\cite{Campos:2018ahf} and references therein.

%% file: tables/tab_ensembles.tex
\begin{tabular}{ccccccccccc}
\toprule
$\beta$ & $\frac{a}{\mathrm{fm}}$ & id & $\frac{L}{a}$ & $\frac{T}{a}$ & $\kappa_\mathrm{l}$ & $\kappa_\mathrm{s}$ & $\frac{m_\pi}{\mathrm{MeV}}$ & $\frac{m_\mathrm{K}}{\mathrm{MeV}}$ & $N_\mathrm{MDU}$ & $\frac{\tau_\mathrm{exp}}{\mathrm{MDU}}$ \\
\midrule
$3.40$ & $0.087$ & H101 & $32$ & $ 96$ & $0.13675962$ & $\kappa_\mathrm{l}$ & $416$ & $416$ & 8064 & $19.9$\\
       &         & H102 & $32$ & $ 96$ & $0.13686500$ & $0.13654934$ & $351$ & $435$ & 8020 & $20.1$\\
       &         & H105 & $32$ & $ 96$ & $0.13697000$ & $0.13634079$ & $276$ & $460$ & 6132 & $20.2$\\
       &         & C101 & $48$ & $ 96$ & $0.13703000$ & $0.13622204$ & $222$ & $470$ & 4220 & $20.4$\\
\midrule
$3.46$ & $0.076$ & H400 & $32$ & $ 96$ & $0.13688848$ & $\kappa_\mathrm{l}$ & $421$ & $421$ & 4180 & $25.4$\\
\midrule
$3.55$ & $0.064$ & H200 & $32$ & $ 96$ & $0.13700000$ & $\kappa_\mathrm{l}$ & $418$ & $418$ & 8000 & $36.0$\\
       &         & N202 & $48$ & $128$ & $0.13700000$ & $\kappa_\mathrm{l}$ & $411$ & $411$ & 3596 & $36.0$\\
       &         & N203 & $48$ & $128$ & $0.13708000$ & $0.13684028$ & $344$ & $441$ & 6172 & $36.0$\\
       &         & N200 & $48$ & $128$ & $0.13714000$ & $0.13672086$ & $283$ & $461$ & 6176 & $36.1$\\
       &         & D200 & $64$ & $128$ & $0.13720000$ & $0.13660175$ & $198$ & $479$ & 4764 & $36.2$\\
\midrule
$3.70$ & $0.050$ & N300 & $48$ & $128$ & $0.13700000$ & $\kappa_\mathrm{l}$ & $418$ & $418$ & 8188 & $59.9$\\
       &         & N302 & $48$ & $128$ & $0.13706400$ & $0.13687218$ & $345$ & $451$ & 8804 & $59.8$\\
       &         & J303 & $64$ & $192$ & $0.13712300$ & $0.13675466$ & $256$ & $473$ & 8584 & $60.3$\\
\midrule
$3.85$ & $0.039$ & J500 & $64$ & $192$ & $0.13685200$ & $\kappa_\mathrm{l}$ & $409$ & $409$ & 6312 & $98.5$\\
       &         & J501 & $64$ & $192$ & $0.13690320$ & $0.13674971$ & $336$ & $446$ & 6540 & $97.2$\\
\bottomrule
\end{tabular}

%% file: tables/tab_kappa_h.tex
\begin{tabular}{cll}
\toprule
$\beta$ & $\kappa_{\mathrm{h}_1}$ & $\kappa_{\mathrm{h}_2}$  \\
\midrule
$3.40$ & $0.123147$    & $0.124056$ \\
$3.46$ & $0.125563292$ & $0.126983423$ \\
$3.55$ & $0.1274374$   & $0.128651119$ \\
$3.70$ & $0.13018588$  & $0.13062697$ \\
$3.85$ & $0.13206693$  & $0.13242984$ \\ 
\bottomrule
\end{tabular}

%% file: 04_extrapolations.tex
\section{Chiral-continuum extrapolations} \label{s:chiralcontinuum}
After having extracted the renormalised quark masses as well as the relevant meson masses from every gauge field configuration ensemble, we proceed with a combined chiral and continuum limit extrapolation in order to obtain the value of the charm quark mass at the physical point, defined by $\phi_2=\phi_2^\mathrm{{phys}}$ and zero lattice spacing. All gauge field ensembles listed in Table~\ref{t:measurements} are included in the analysis, except for H200 with a spatial extent of $\sim 2\,$fm that we consider too small.\footnote{Finite-volume effects for light current quark masses and the combined meson masses $\phi_2$ and $\phi_4$ on our set of ensembles were investigated in Refs.~\cite{Bruno:2016plf, Bruno:2019vup} via a comparison of the results from the ensembles H200 and N202. They were found to be significantly smaller than the available statistical accuracy. We observe the same to be true for the charmed observables studied here. In general, current quark masses and particularly masses of heavy mesons~\cite{Colangelo:2010ba} are expected to be unaffected by moderate violations of the conditions that typically qualify lattice QCD ensembles as resembling an infinite volume situation.} In the extrapolation, we include our different definitions (\ref{e:charm_m_cc}), (\ref{e:charm_m_lc}) and (\ref{e:charm_m_rd}) of the renormalised charm quark mass, in conjunction with eq.~(\ref{e:M_over_m_nf3}) and made dimensionless by attaching the factor $\sqrt{8t_0}$. For the renormalisation constant $Z$, as well as the improvement coefficients $b_\mathrm{A}-b_\mathrm{P}$ and $b_\mathrm{m}$ entering these definitions, the numbers denoted as `LCP-1' in Ref.~\cite{deDivitiis:2019xla} are inserted, as their determination was specifically suited for the heavier quark regime and provides high statistical accuracy. As consistency checks, we also tried using the `LCP-0' values of these parameters from Ref.~\cite{deDivitiis:2019xla}, as well as those from the alternative Ward identity determination of $Z$ in Ref.~\cite{Heitger:2020mkp}, and found slight differences at finite lattice spacing but full agreement in the continuum limit. 
We utilise the results for $r_\mathrm{m}$ from Ref.~\cite{Heitger:2021bmg} in the corresponding terms of $\mathrm{O}(aM_\mathrm{sum})$ in eqs.~(\ref{e:mpcac_renormalised_rm2}) and (\ref{e:bmw_improved_difference}).\footnote{As argued at the end of Section~\ref{s:quarkmassesonthelattice}, the terms proportional to $\bar{b}_\mathrm{m}$ and ($\bar{b}_\mathrm{A}-\bar{b}_\mathrm{P}$) are not expected to be significant and therefore ignored in our improvement procedure.}However, our data reveals that neglecting these terms only has an insignificant effect on the final result. 

Eventually, after the joint extrapolation of $\sqrt{8t_0}M_{\mathrm{c}}$ to the physical point, the final value of the charm quark mass in physical units then simply stems from division by $\sqrt{8t_0^\mathrm{phys}}$.

For the charm quark mass in the continuum we assume the functional form
\begin{align}
\sqrt{8t_0}M^\mathrm{continuum}_{\mathrm{c}}(\phi_2,\phi_\mathrm{H},0)=c_0+c_\mathrm{L} \phi_2+ c_\mathrm{c} \phi_\mathrm{H}\,, \label{e:chiral_part}
\end{align}
which is guided by the following reasoning:
Having corrected the chiral trajectory by means of the mass derivatives in eq.~(\ref{e:shifted_observable}), the light quark mass dependence is governed by the sea pion mass only, whereas the kaon mass is implicitly fixed by the $\mathrm{Tr}[M_q]=\mathrm{const.}$ condition realised along the trajectory in lattice parameter space towards the physical point. Based on the Gell-Mann-Oakes-Renner relation \cite{GellMann:1968rz}, we thus assume the leading contribution to be proportional to $\phi_2$.
Moreover, we model the interpolation to the physical value of the bare charm quark mass via
\begin{align}
\phi_\mathrm{H}=\sqrt{8t_0}m_\mathrm{H}\,,
\end{align}
where $m_\mathrm{H}$ labels the mass of a charmed meson for which we either chose the flavour averaged D meson mass, $m_{\bar{\mathrm{D}}}$, or the mass of the pseudoscalar charmonium meson, $m_{\eta_\mathrm{c}}$ (neglecting all quark-disconnected contributions).
Motivated by heavy quark effective theory, we expect the charm quark mass to be linearly dependent on the meson mass of choice \cite{Neubert:1993mb}.

For the continuum part we presume the leading cutoff effects to be of $\mathrm{O}(a^2)$ as all relevant renormalisation constants and improvement coefficients that propagate into our computation are known non-perturbatively. In addition, we also allow for terms describing the next two higher orders, $\mathrm{O}(a^3)$ and $\mathrm{O}(a^4)$. These cutoff effects may be quark mass independent or explicitly depend on the quark masses. Therefore, our general ansatz for the lattice spacing dependence of the charm quark mass is parametrised by
\begin{align}
c_{M_{\mathrm{c}}}(\phi_2,\phi_\mathrm{H},a)=\frac{a^2}{8t_0}\bigg(c_1 + c_2\phi_\mathrm{H}^2\bigg)+\frac{a^3}{(8t_0)^{3/2}}\bigg(c_3 + c_4\phi_\mathrm{H}^3\bigg)+\frac{a^4}{(8t_0)^{2}}\bigg(c_5 + c_6\phi_\mathrm{H}^4\bigg)\,; \label{e:continuum_part}
\end{align}
all cutoff effects proportional to the light quark masses are already neglected here, because it turned out that these effects cannot be resolved in the fits. Furthermore, for now we also neglect terms proportional to mixed powers of $\tfrac{a}{\sqrt{8t_0}}$ and $\phi_\mathrm{H}$, such as $\tfrac{a^3}{(8t_0)^{3/2}}\phi_\mathrm{H}^2$ or $\tfrac{a^4}{(8t_0)^2}\phi_\mathrm{H}^2$. We will demonstrate below that the inclusion of these terms (which a priori cannot be excluded on theoretical grounds) has no effect on our final result. As explained at the end of Section \ref{s:quarkmassesonthelattice}, we do not expect terms of $\mathrm{O}(g_0^4a \mathrm{Tr}\left[M_\mathrm{q}\right])$ to contribute significantly to the cutoff effects.  We have checked explicitly that the inclusion of such terms does not alter the continuum limit and found the corresponding fit parameters to vanish within errors.

To arrive at a combined model for the global fitting procedure, we can compose the chiral and continuum parts either linearly, by adding eqs.~(\ref{e:chiral_part}) and~(\ref{e:continuum_part}), viz.
\begin{align}
\sqrt{8t_0}M^\mathrm{linear}_{\mathrm{c}}(\phi_2,\phi_\mathrm{H},a)=\sqrt{8t_0}M^\mathrm{continuum}_{\mathrm{c}}(\phi_2,\phi_\mathrm{H},0) + c_{M_{\mathrm{c}}}(\phi_2,\phi_\mathrm{H},a)\,,
\end{align}
or in a non-linear fashion by multiplication of the two:
\begin{align}
\sqrt{8t_0}M^\mathrm{non-linear}_{\mathrm{c}}(\phi_2,\phi_\mathrm{H},a)=\sqrt{8t_0}M^\mathrm{continuum}_{\mathrm{c}}(\phi_2,\phi_\mathrm{H},0) \times (1 + c_{M_{\mathrm{c}}}(\phi_2,\phi_\mathrm{H},a))\,;
\end{align}
the $c_i$ in (\ref{e:chiral_part}) and~(\ref{e:continuum_part}) are fit parameters.
From now on we will refer to different model parametrisations with the following labelling scheme:
\begin{itemize}
	\item The first entry is either 'l' for the linear combination of the chiral and continuum parts or 'nl' for the non-linear variant.
	\item The second entry indicates the term which accounts for cutoff effects proportional to $a^2$. Here, '1' stands for the term parametrised by $c_1$, whereas '2' stands for the (heavy meson) mass dependent one, parametrised by $c_2$. When both pieces, proportional to $c_1$ and $c_2$, are included, the entry is labelled by '3'.
	\item The third entry represents the contribution accounting for cutoff effects of $\mathrm{O}(a^3)$. Here, '0' specifies fits without any contribution of this order, whereas '1' does so for fits with the $c_3$-term, '2' for fits with the $c_4$-term and '3' for fits with both of them.
	\item The fourth entry labels the terms describing discretisation effects of $\mathrm{O}(a^4)$. In analogy to the third entry, '0' labels no term of this order, '1' the term proportional to $c_5$, '2' the one proportional to $c_6$ and '3' both of them.
	\item Finally, the fifth entry indicates whether a pion mass dependence is included in the fit, i.e., '0' marks models without and '1' models with the term parametrised by the coefficient $c_\mathrm{L}$.
\end{itemize}
As an example, the model labelled by ('l', 3, 0, 1, 0) corresponds to the fit function
\begin{align}
\sqrt{8t_0}M^\mathrm{('l', 3, 0, 1, 0)}_\mathrm{c}(\phi_2,\phi_\mathrm{H},a)=c_0+ c_\mathrm{c} \phi_\mathrm{H}+a^2\bigg(\frac{c_1}{8t_0} + c_2m_\mathrm{H}^2\bigg)+a^4\bigg(\frac{c_5}{(8t_0)^{2}}\bigg)\,.
\end{align}
We always include at least one term describing cutoff effects of $\mathrm{O}(a^2)$ and restrict the maximum number of terms related to the cutoff effects to three. In total this amounts to $K=108$ different models, with which we attempt to fit our data.

In order to estimate the optimal parameters for a given model (generically designated $p_i$ and $f$ in the formula below), we employ an extended $\chi^2$ minimisation procedure, taking into account the errors of both, the independent and the dependent variables \cite{Boggs1989}, viz.
\begin{align}
\chi^2=\big(f(p_i;\tilde{x}_a)-y_a, \tilde{x}_a-x_a\big)C^{-1}\big(f(p_i;\tilde{x}_a)-y_a, \tilde{x}_a-x_a\big)^\mathrm{T}\,.
\end{align}
The independent variables $x_a$ are promoted to fit parameters $\tilde{x}_a$, which albeit stay constrained to their initial values $x_a$ and thereby only indirectly affect the number of degrees of freedom.
In our case, the full correlation matrix is ill-conditioned and eludes safe inversion.\footnote{Whereas Monte Carlo data from different gauge field ensembles is uncorrelated by construction, the full covariance matrix does not assume a block diagonal form, but receives additional contributions induced by common renormalisation and improvement parameters as well as the mass shifting procedure.}
For this reason we ignore all correlations and set the off-diagonal elements of the error covariance matrix $C$ explicitly to zero.

While minimising the uncorrelated $\chi^2$ yields reliable maximum likelihood estimators for our parameters (see Ref.~\cite{PhysRevD.49.2616}), the minimal value of the uncorrelated $\chi^2$ cannot serve as a meaningful indicator for the goodness of fit, as we do not know the true number of degrees of freedom. In practice, we would thus obtain values of $\chi^2/(N-k)<1$ also for models that do not describe the data well.
We circumvent this issue by evaluating $\chi^2_\mathrm{exp}$, the so-called `expected' $\chi^2$ proposed and advocated for use in such situations in Ref.~\cite{Bruno:2020xxxx},\footnote{Note that $\chi^2_\mathrm{exp}$ was already successfully applied before in a fit analysis of HQET correlators in the context of extracting semi-leptonic form factors~\cite{Bahr:2019eom}.}which equals the expectation value of $\chi^2$ given normally distributed data with the full covariance matrix $C$. As explained in this reference, it qualifies to derive a corrected measure for the goodness of fit,
\begin{align}
\chi^2_\mathrm{corrected}=\frac{(N-k)}{\chi^2_\mathrm{exp}}\chi^2\,, \label{e:corrected_chisq}
\end{align}
in the sense that $\langle \chi^2_\mathrm{corrected} \rangle / (N-k)=1$ when the model describes the data.

\subsection{Systematic effects} \label{ss:systematics}
In order to address systematic effects in our computation of the charm quark mass quantitatively, we classify our data in different categories w.r.t. the various choices and technical ingredients that characterise the analysis:
\begin{itemize}
	\item As estimator for the charm quark mass on the lattice, we pick the definitions $\overline{m}_\mathrm{R, c}$, $m_\mathrm{R, c}^{(\mathrm{c})}$ or $m_\mathrm{R, c}^{(\mathrm{rd})}$ introduced in eqs.~(\ref{e:charm_m_cc}), (\ref{e:charm_m_lc}) and (\ref{e:charm_m_rd}).
	\item For the bare current quark masses involved, either standard derivatives according to eq.~(\ref{e:standard_derivatives}) or its improved version in eq.~(\ref{e:improved_derivatives}) are used.
	\item To fix the physical charm quark mass, we either employ the flavour-averaged D meson mass, $m_{\bar{\mathrm{D}}}$, or the mass of the pseudoscalar charmonium groundstate, $m_{\eta_\mathrm{c}}$ (neglecting the quark-disconnected contributions).
	\item To ensure that cutoff effects are properly described by the fits, we either include all gauge field ensembles, i.e., covering a range of lattice spacings $0.039\leq \frac{a}{\mathrm{fm}} \leq 0.087$, or a cut of $0.039\leq \frac{a}{\mathrm{fm}} \leq 0.076$ is imposed on the data, i.e., all ensembles at $\beta=3.4$ are excluded.
\end{itemize}
\begin{figure}
	\centering
	\includegraphics[width=\linewidth]{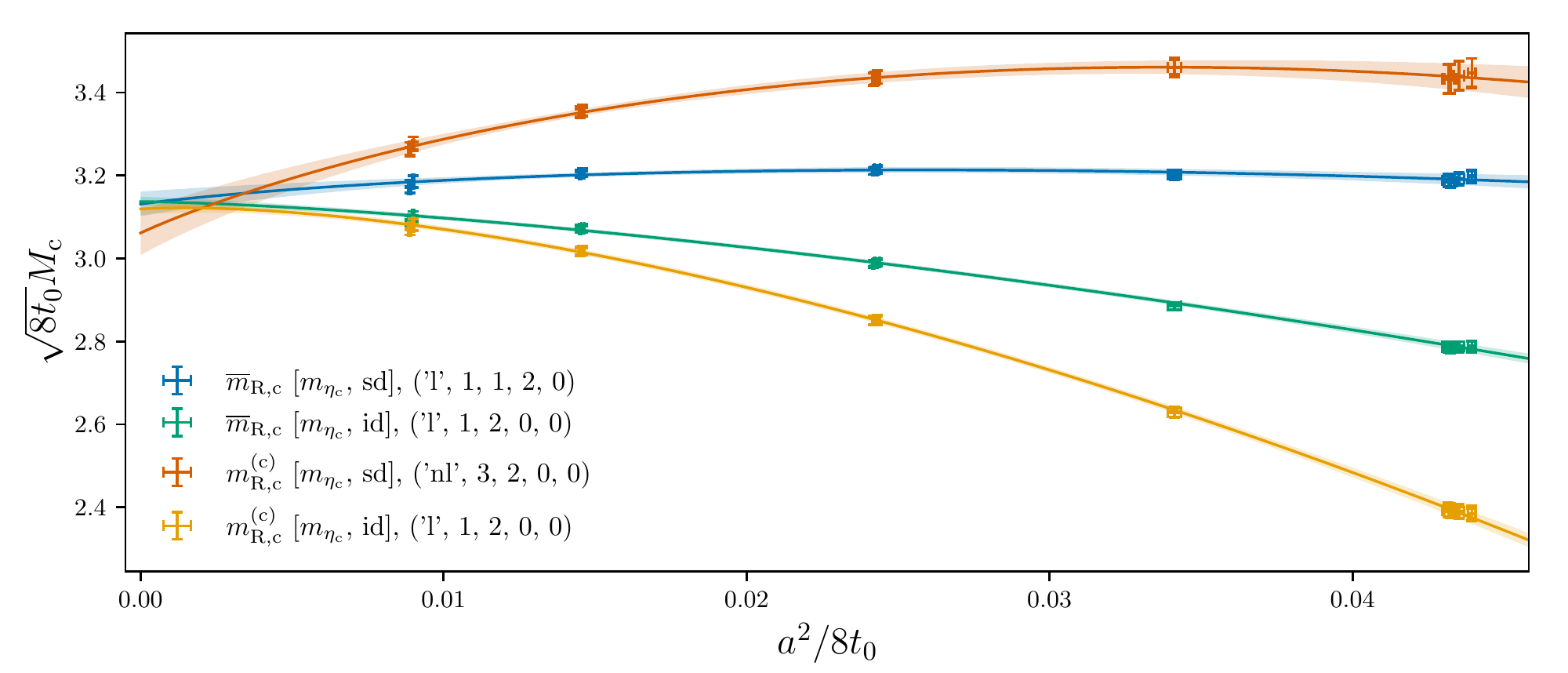}
	\includegraphics[width=\linewidth]{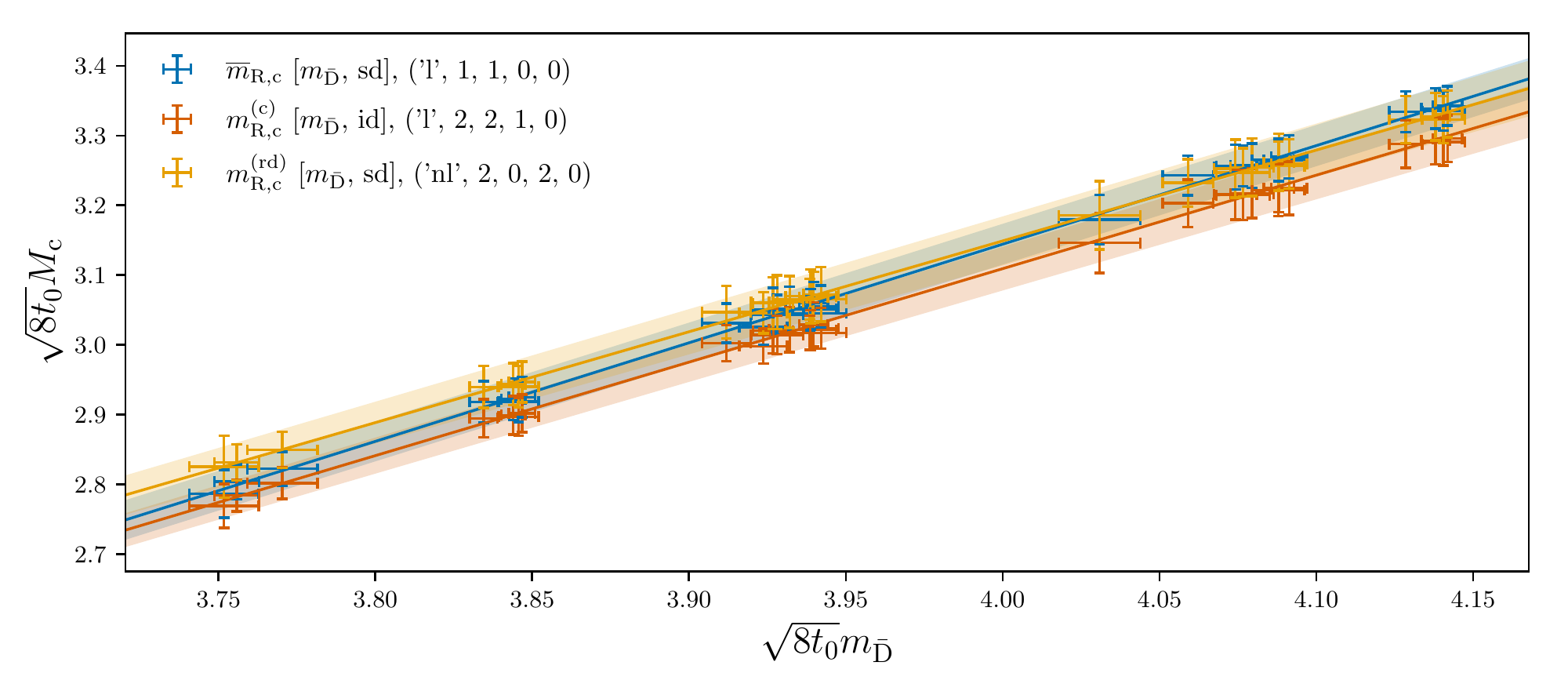}
	\includegraphics[width=\linewidth]{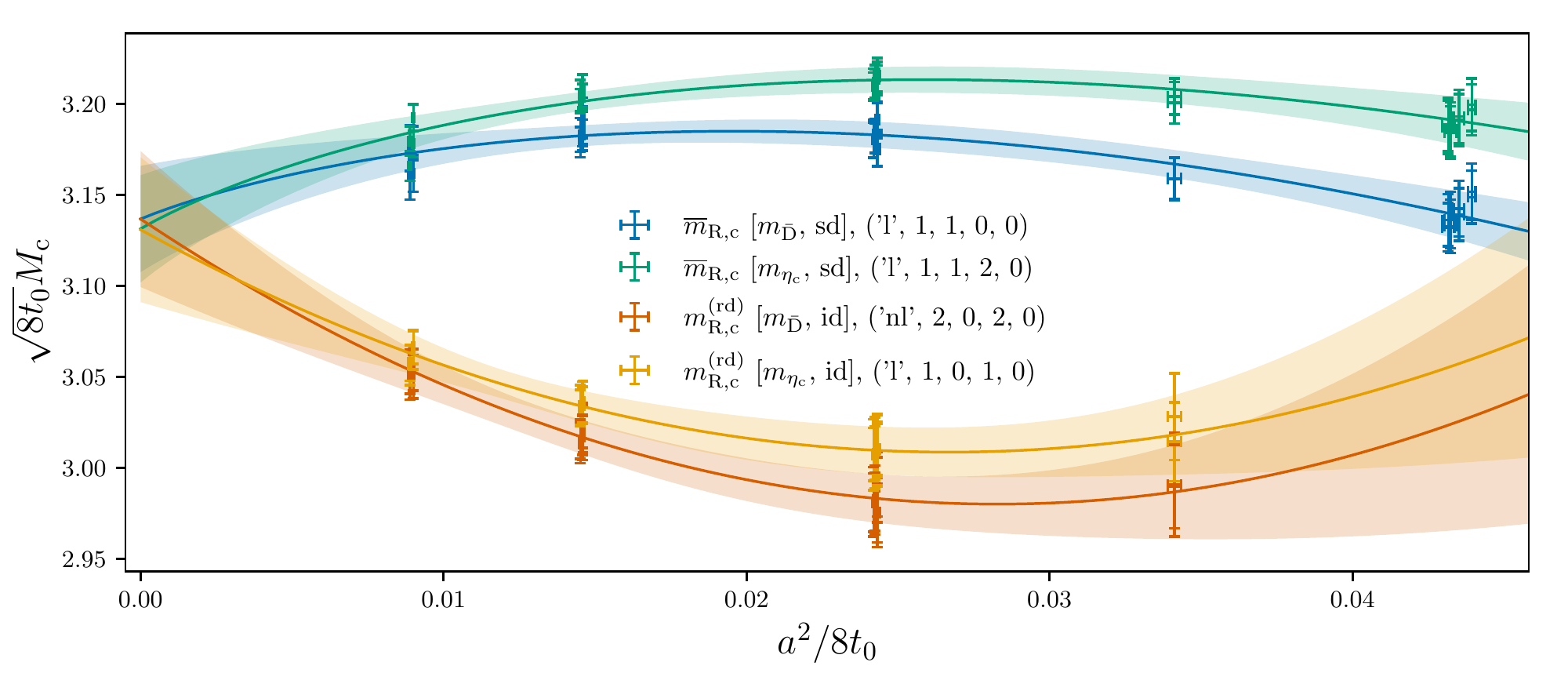}
	\caption{Comparison of best fits according to the AIC in different categories. To map out two-dimensional representations of the fits, other dependencies than the ones illustrated in the plots have been projected to the physical point. The top panel reproduces the influence of utilising standard versus improved lattice derivatives to extract bare PCAC quark masses, while the middle one exhibits the clearly linear dependence of the charm quark mass estimators on the dimensionless heavy meson mass, $\phi_\mathrm{H}=\sqrt{8t_0}m_\mathrm{H}$, in case of $m_\mathrm{H}=m_{\bar{\mathrm{D}}}$. The bottom panel reflects the sensitivity of the joint global extrapolation results to the choice of $\phi_\mathrm{H}$ for fixing the physical charm point.   
    The labelling convention as well as further details are given in the text.}
	\label{fig:comp}
\end{figure}
This amounts to 24 categories in total. For each category we perform fits with all 108 fit forms specified in the previous subsection, resulting in a total of 2592 fits.
In the following we discuss some of these fits in more detail. As measure for the best fit in each category we adopt the Akaike information criterion (AIC) \cite{Akaike1998}
based on the above-introduced $\chi^2_\mathrm{corrected}$,
\begin{align}
\mathrm{AIC}=\chi_\mathrm{corrected}^2+2k\,,
\end{align}
where $k$ is the number of parameters in the fit function.

In Figure~\ref{fig:comp} we compare the best fits in different representative categories.
The top part illustrates the effect of using improved derivatives in the PCAC masses on the definitions of the charm quark mass. Starting with $\overline{m}_{\mathrm{R},\mathrm{c}}$, we see that improvement of the derivative leads to larger cutoff effects compared to the standard one. However, both variants agree very well in the continuum limit. This is also the case for $m_{\mathrm{R},\mathrm{c}}^{(\mathrm{rd})}$ (not displayed in the figure), but not so for $m_{\mathrm{R},\mathrm{c}}^{(\mathrm{c})}$ which is also shown in the top part of Figure~\ref{fig:comp}. Here, the variant with improved derivative coincides with the other results, while the extraction based on the standard derivative tends to smaller values. Moreover, besides statistical uncertainties being larger for the coarser lattice spacings, cutoff effects are expected to be generically larger for doubly charmed PCAC masses. For these reasons we decide to exclude all determinations through $m_{\mathrm{R},\mathrm{c}}^{(\mathrm{c})}$ with the standard derivative in the final analysis.

Regarding the lattice spacing dependence, we observe for all categories in our analysis that terms of $\mathrm{O}(a^3)$ or $\mathrm{O}(a^4)$ have to be taken into account when considering ensembles with $a>0.06\,$fm. For finer resolutions, terms modelling $a^2$-contributions are sufficient to describe the data reliably. The quadratic scaling regime therefore starts later than one could have naively expected.\footnote{A similar effect has been observed earlier in a computation of the $D_\mathrm{s}$ meson decay constant in the quenched approximation~\cite{Heitger:2008jq}.}Due to the purely non-perturbative inputs of renormalisation and improvement factors, lattice artefacts of $\mathrm{O}(a)$ are absent.

For all variants that we consider, fixing the physical value of the charm quark mass works completely satisfactory with a term linear in the (heavy) meson mass $\phi_\mathrm{H}=\sqrt{8t_0}m_\mathrm{H}$. This is demonstrated in the middle part of Figure~\ref{fig:comp}, which depicts the dependence of the three different charm quark mass definitions on $m_\mathrm{H}=m_{\bar{\mathrm{D}}}$. Employing $m_\mathrm{H}=m_{\eta_\mathrm{c}}$ instead to fix the physical charm quark mass, we find a very similar behaviour. Note that for $\beta=3.85$ the two values chosen for $\kappa_\mathrm{c}$ do not enclose the physical charm point; so in this case we cannot interpolate, but have to rely on a slight extrapolation to the target. However, the perfectly linear behaviour reflected by the data suggests that this procedure is entirely legitimate.

The bottom part of Figure~\ref{fig:comp} displays how our data and the respective joint chiral and continuum extrapolations depend on $\phi_\mathrm{H}=\sqrt{8t_0}m_\mathrm{H}$ that is chosen to fix the bare charm quark mass to its physical value. Compared to $m_\mathrm{H}=m_{\bar{\mathrm{D}}}$, the variants based on $m_{\eta_\mathrm{c}}$ in general tend to slightly larger values at finite lattice spacing. However, this ambiguity vanishes in the continuum limit as expected.

Since the fit distinguished as the best one by the AIC (or any other sensible measure) could in fact be also an outlier, we opt for performing a model averaging procedure within each category, similar to what has been proposed in Ref.~\cite{Jay:2020jkz}.
In analogy to this reference, we define the model average via
\begin{align}
\langle M_\mathrm{c}\rangle = \sum_{k=1}^{K}w_k\langle M_\mathrm{c}\rangle_k\,, \label{e:model_average}
\end{align}
with the respective model weights $w_k$ and $k$ indexing the alltogether $K=108$ models (i.e., fit forms) within each of the 24 categories at disposal. For the systematic error arising from this model selection procedure we assume
\begin{align}
\sigma^2_{M_\mathrm{c}}=\sum_{k=1}^{K}w_k\langle M_\mathrm{c}\rangle_k^2-\Bigg( \sum_{k=1}^{K}w_k\langle M_\mathrm{c}\rangle_k \Bigg)^2\,.
\label{e:model_average_error}
\end{align}
For uniform weights, $w_k\equiv 1/K$, this expression reduces to the variance of the different model estimates. Yet instead of setting the weights uniformly, we rather set the $w_k$ for each fit relying on the AIC,
\begin{align}
w_k^\mathrm{AIC}=N\exp\bigg(-\frac{1}{2}\mathrm{AIC}\bigg)\,, 
\end{align}
where the normalisation constant $N$ is chosen to ensure $\sum_{k=1}^{K}w_k^\mathrm{AIC}=1$.

\begin{figure}
	\centering
	\includegraphics[width=\linewidth, trim={0 6.5pt 0 0},clip]{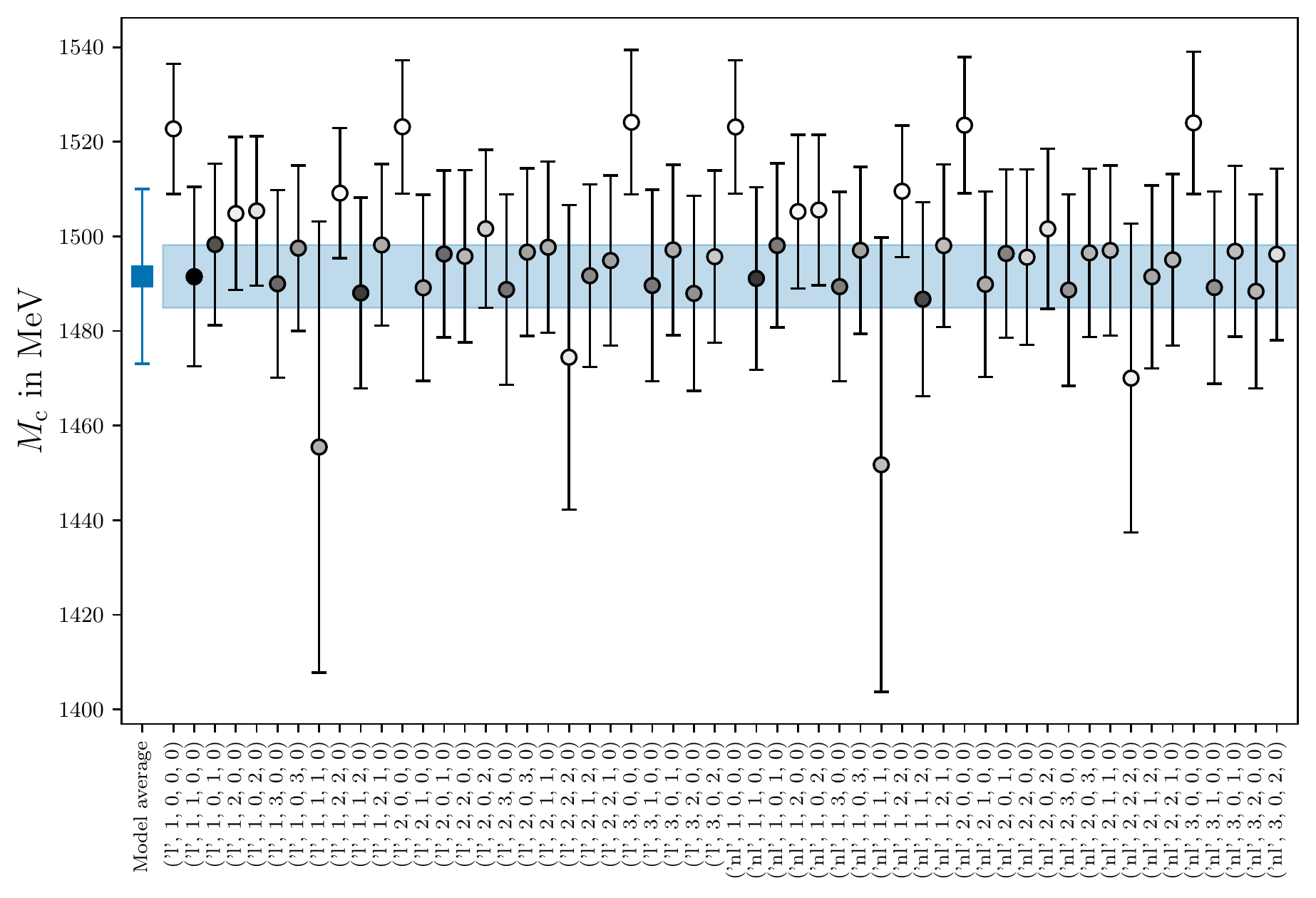}
	\includegraphics[width=\linewidth, trim={0 0 0 6.5pt},clip]{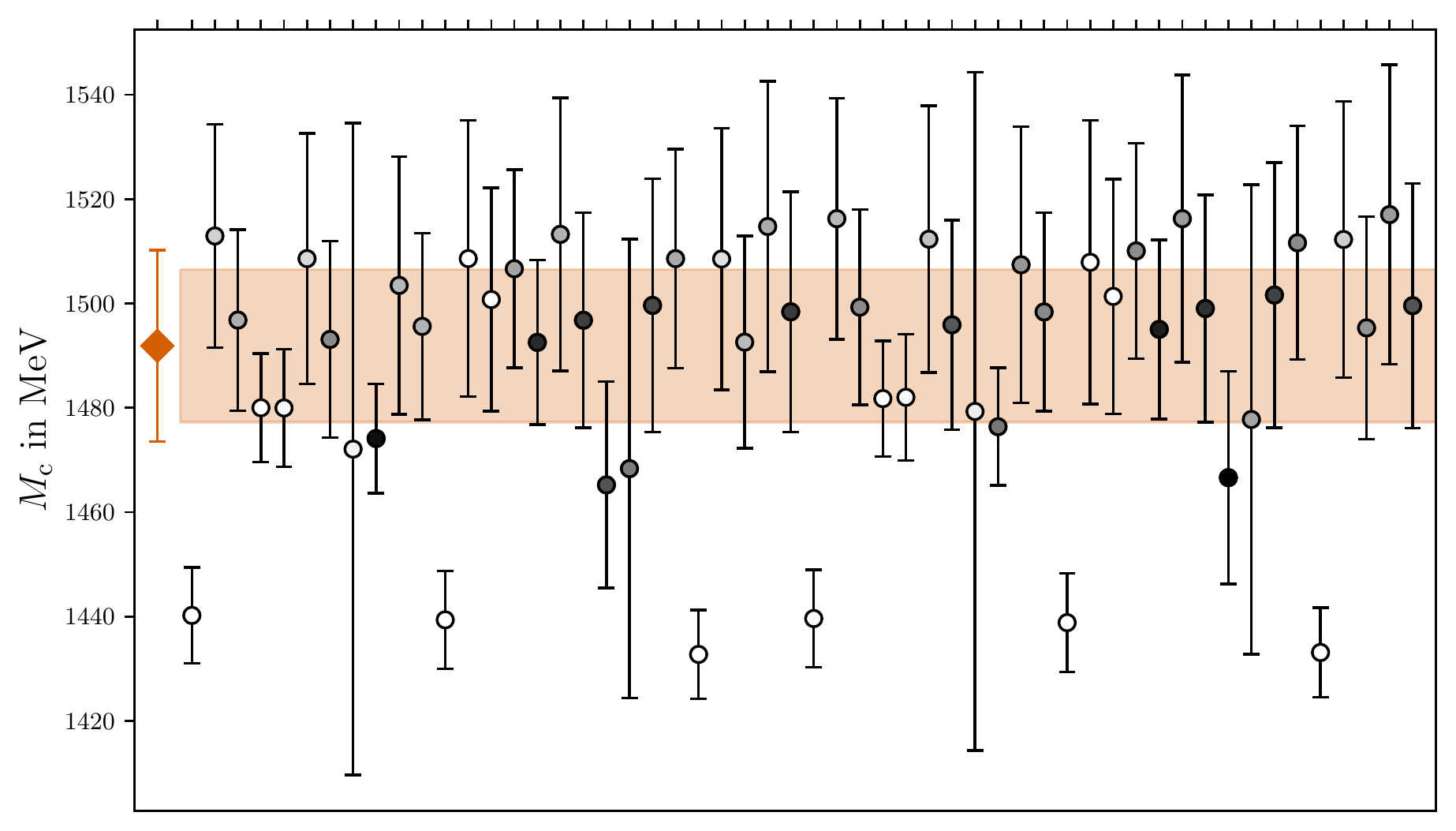}
	\caption{Model averaging procedure for two representative categories: $\overline{m}_{\mathrm{R},\mathrm{c}}$ with $m_\mathrm{H}=m_{\bar{\mathrm{D}}}$, standard lattice derivative and all lattice spacings included (top panel); and $m^{(\mathrm{rd})}_{\mathrm{R},\mathrm{c}}$ with $m_\mathrm{H}=m_{\eta_\mathrm{c}}$, improved derivative and all lattice spacings included (bottom panel). Each circle corresponds to the result of the model indicated along the horizontal axis. The opacity of the circles is proportional to their weights contributing to the average. The filled square and diamond give the respective model averages with statistical error according to eq.~(\ref{e:model_average}), while the transparent bands indicate the systematic errors arising from the modeling procedure according to eq.~(\ref{e:model_average_error}), both of which are added in quadrature. Fits with pion mass dependence are not shown, as their contribution to the model averages is subdominant. Note that the model averages obtained as visualised here correspond to one data point each in the subsequent Figure~\ref{fig:mcharmsystematics}.}
	\label{fig:model_averaging}
\end{figure}
\begin{figure}
	\centering
	\includegraphics[width=\linewidth]{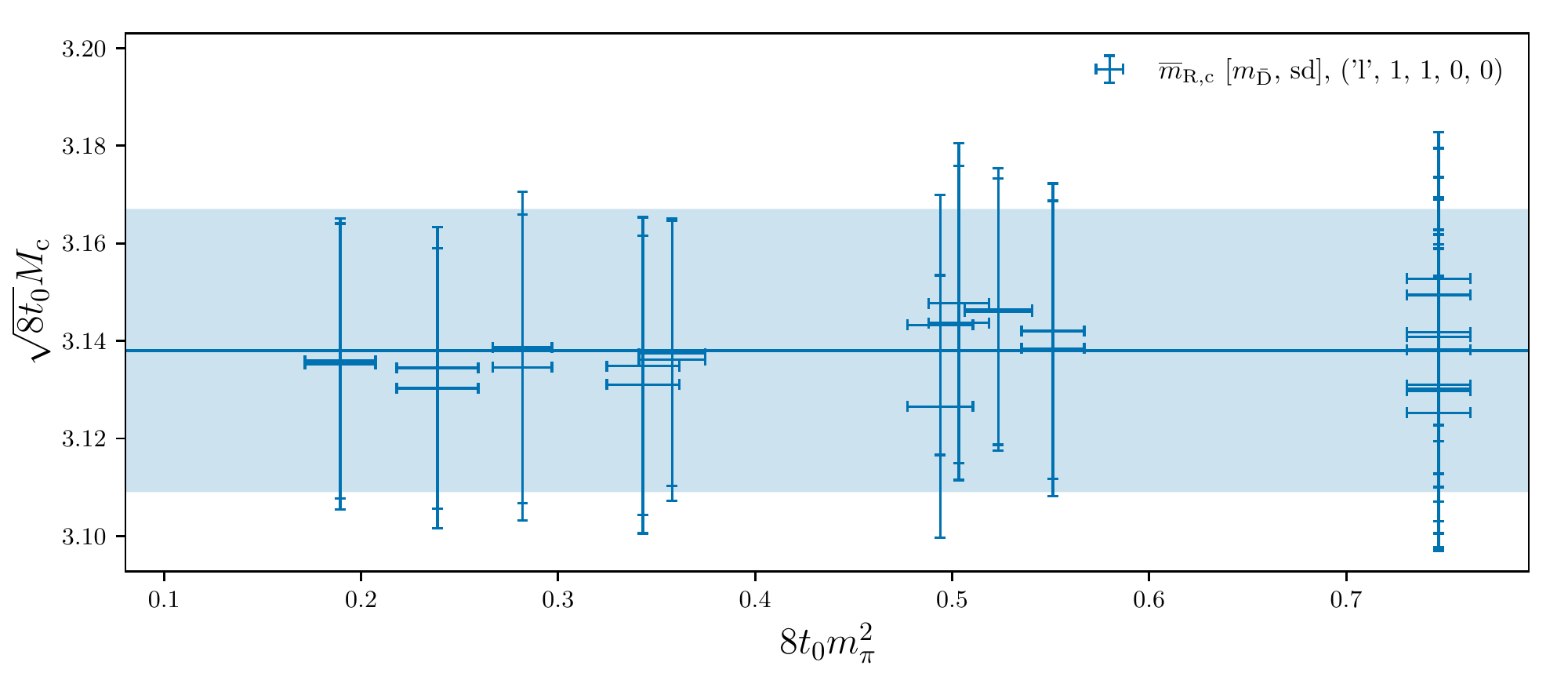}
	\caption{Pion mass dependence and best chiral fit according to the AIC for the charm quark mass as defined in eq.~(\ref{e:charm_m_lc}). Note that $\phi_4$ is held constant along the chiral trajectory. To map out a two-dimensional representation of the fit, other dependencies than the one illustrated in the plot have been projected to the physical point.  
	The labelling convention is given in the text.}
	\label{fig:comp_mpi}
\end{figure}

In Figure~\ref{fig:model_averaging} the outcome of the model averaging procedure is summarised for two representative examples among the 24 categories. The individual circles correspond to the result of fitting the data to the respective model, labelled along the horizontal axis according to scheme introduced above. Their opacity indicates the associated weight in the average, the most opaque points having largest weight. The filled square and diamond mark the model averages with their statistical errors, while the shaded bands depict their systematic uncertainties emerging from the model averaging procedure. As a consequence of the definitions of the renormalised charm quark mass, eqs.~(\ref{e:charm_m_cc}), (\ref{e:charm_m_lc}) and (\ref{e:charm_m_rd}), which were suitably adapted to the position of the CLS ensembles in physical parameter space and the approach towards the chiral limit ensuing from it, we expect the (sea) pion mass dependence of $M_\mathrm{c}$ to be small. In fact, the best fits in all 24 categories (and also the majority of fits with a relevant weight) do not incorporate a pion mass dependence. As a representative example we illustrate the pion mass dependence for the charm quark mass defined in eq.~(\ref{e:charm_m_lc}) in Figure~\ref{fig:comp_mpi}. Other quark mass definitions show the same flat behaviour along the chosen chiral trajectory for which $\phi_4$ is kept constant.

As can be inferred from the top part of Figure~\ref{fig:model_averaging}, almost all fits with non-negligible weight agree very well, which is reflected by the small systematic error. In this case, singling out only the best fit would give nearly an identical result compared to the model average.
In the bottom part of Figure~\ref{fig:model_averaging}, the best fit is ('nl', 2, 1, 1, 0) and differs from the final model average by $\sim 1\sigma$. However, owing to the very construction of this average, the large number of models that describe the data reasonably, outweigh the few outliers. The somewhat wider spread of the fits with more significant weight is revealed by the larger systematic model averaging error in this case.

\begin{figure}[t]
	\centering
	\includegraphics[width=\linewidth]{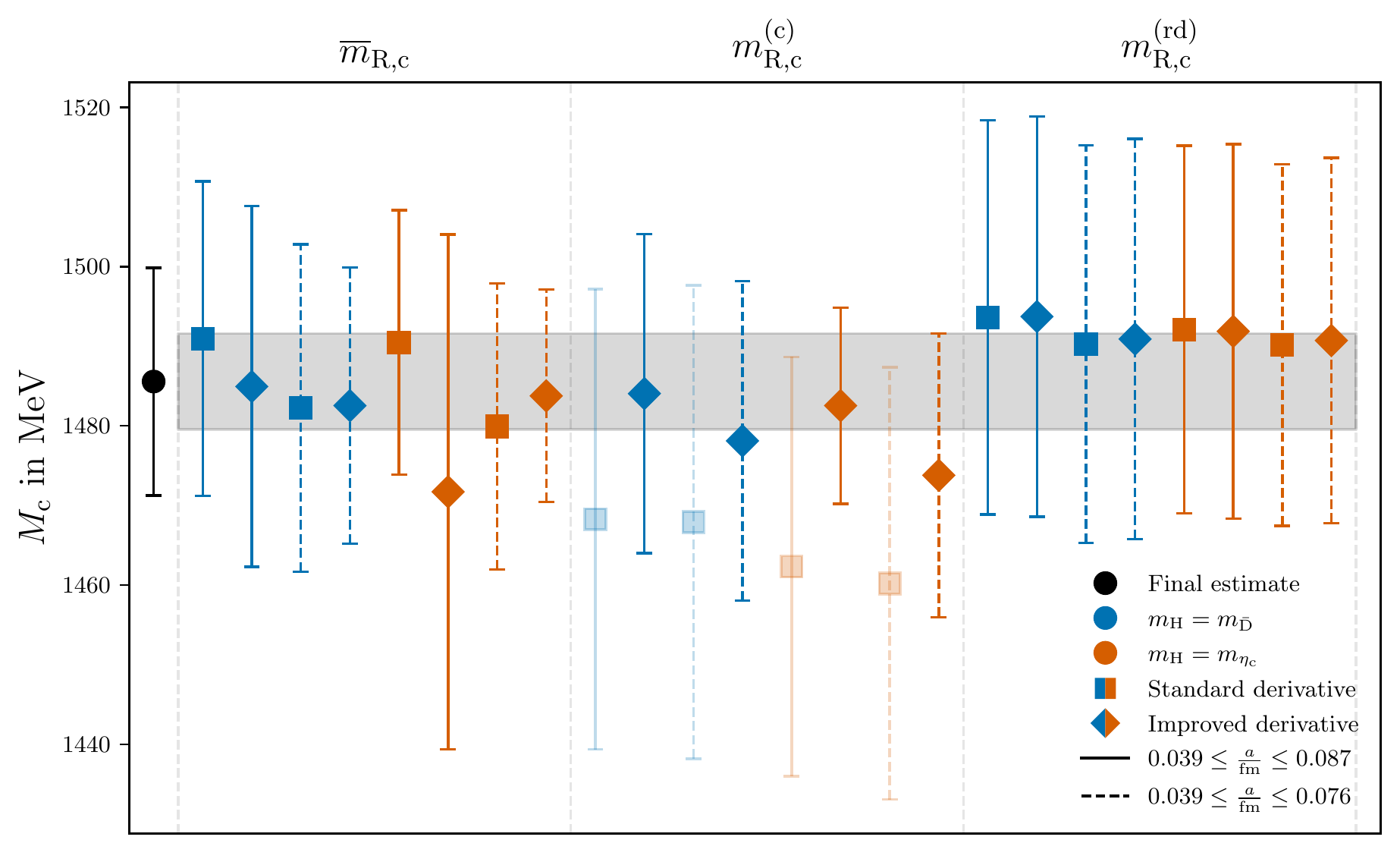}
	\caption{Model averages in each category as evaluated through the AIC weighting prescription. Results for the $m_{\mathrm{R},\mathrm{c}}^{(\mathrm{c})}$ variant with the standard derivative, depicted as transparent points, are excluded from our final estimates as explained in the text. The shaded region indicates the estimate of the systematic error obtained by the weighted standard deviation of the model averages over all categories. (The legend recalls the key characteristics of every category.)}
	\label{fig:mcharmsystematics}
\end{figure}

Figure~\ref{fig:mcharmsystematics} displays the mean values and errors from the model averages within each of the 24 categories specified above.
In principle, the result in every single category represents a valid determination of the charm quark mass, which means that for infinite statistics we would expect all 24 determinations to coincide perfectly. Hence, to decrease the final error, one could even consider combined fits of different categories.
However, as we will discuss in the next section, our error is not dominated by the statistical error arising from the underlying Monte Carlo data, but by the propagated error from external inputs. For this reason we have decided to conservatively take the spread of the individual categories' results as additional measure for the systematic error of our calculation.

For our final result of $M_{\mathrm{c}}$ we chose the weighted average over all categories in Figure~\ref{fig:mcharmsystematics}, except (as argued before) for the ones with $m_{\mathrm{R},\mathrm{c}}^{(\mathrm{c})}$ and the standard derivative. The weights are given by the inverse of the respective errors (i.e., statistical and systematic ones added in quadrature). From the weighted standard deviation (\ref{e:model_average_error}) of the model averages within these categories we estimate the size of the systematic error in our determination of the charm quark mass. However, we have convinced ourselves that an unweighted average leads to a number which is indistinguishable from the weighted average in relation to the total error for the final result quoted below.\footnote{Note that the model average according to eq.~(\ref{e:model_average}) is applied on the level of the samples (available from the $\Gamma$-method in our case) and therefore all correlations are properly taken into account in the computation of the statistical error of $\langle M_\mathrm{c}\rangle$.}

In order to check the robustness of our analysis method, we have also induced different weights in the model averaging procedure, devised to set different penalties on the number of model parameters and thereby to favour over- or underfitting opposed to the standard procedure. Specifically, the Bayesian information criterion \cite{schwarz1978} as well as the corrected AIC \cite{Hurvich1989} were tried, which give higher penalties for models with more fit parameters, while basing the weights on just $\chi_\mathrm{corrected}^2$ introduces no penalty at all. The resulting variation on our final estimate of $M_\mathrm{c}$ when using these different weights is by far smaller than our systematic error as assessed above. Furthermore, as already mentioned earlier, we investigated the effect of allowing for terms such as $\tfrac{a}{\sqrt{8t_0}}(am_\mathrm{H})^2$ or $\tfrac{a^2}{8t_0}(am_\mathrm{H})^2$ in the generic fit ansatz for the joint chiral and continuum limit extrapolation, thus increasing the number of possible models in each category to $K=168$. Again, also this extension of the number of models has a completely negligible effect on the final result. In summary, we therefore conclude that grounding our analysis on a large number of models and categories very consistently yields a reliable result that is fairly independent of human bias, because the manual input is only limited to a theoretically well justified pre-selection of considered categories and models. 

%% file: 05_results.tex
\section{Result} \label{s:results}
Following the data analysis presented in the foregoing sections, we quote for the RGI charm quark mass from our lattice QCD computation as final result
\begin{align}
M_{\mathrm{c}}(N_\mathrm{f}=3)=1486(14)(14)(6)\,\mathrm{MeV}=1486(21)\,\mathrm{MeV}\,,\label{e:3f_rgi}
\end{align}
which constitutes an unambiguous result in (the continuum limit of) the $N_\mathrm{f}=2+1$ theory studied here. The first error is statistical, the second stems from the factor converting renormalised into RGI quark masses, available from Ref.~\cite{Campos:2018ahf}, while the third one represents our estimate of systematic effects.

In Table~\ref{tab:systematic_error} we summarise the error budget of our computation.
\begin{table}
	\setlength{\abovecaptionskip}{10pt}
	\setlength{\belowcaptionskip}{0pt}
	\caption{Error budget of our computation, quantified by the contributions to the squared error of our final estimate for $M_{\mathrm{c}}$. We group them into the three main categories `Statistical error', `Renormalisation group running' and `Systematic uncertainty', which comprise the final (total) error, where the listed individual subcategories indicate how these uncertainties are distributed among the main categories.}
	\label{tab:systematic_error}
	\centering
	\begin{tabular}{lrc}
		\toprule
		\textbf{Statistical error} & \textbf{44.8\% }&\\
		\hspace{18pt}correlation functions && 2.2\% \\
		\hspace{18pt}$\mathrm{O}(a)$ improvement && 14.2\% \\
		\hspace{18pt}renormalisation && 8.7\% \\
		\hspace{18pt}scale setting && 19.7\% \\
		\hspace{18pt}physical meson masses && <$\,$0.1\% \\
		\textbf{Renormalisation group running} & \textbf{45.9\% }&\\
		\textbf{Systematic error} & \textbf{9.3\% }&\\
		\hspace{18pt}model selection && 1.1\% \\
		\hspace{18pt}input data variation && 8.2\% \\
		\bottomrule
	\end{tabular}
\end{table}
The most dominant contribution to the total error originates from the (universal) conversion factor $M/m_\mathrm{R}(\mu_\mathrm{had})$ that connects renormalised quark masses at finite scale via non-perturbative renormalisation group running to their RGI counterparts (cf.~Subsection~\ref{ss:MRGI}) and was obtained for the three-flavour theory in Ref.~\cite{Campos:2018ahf}. The next relevant contributions come from the scale setting of Ref.~\cite{Bruno:2016plf}, and the $\mathrm{O}(a)$ improvement $b$-coefficients multiplying the quark mass dependent pieces, non-perturbatively obtained in Ref.~\cite{deDivitiis:2019xla}. For the systematic uncertainty, which is made up of two contributions, the one from the spread of outcomes of physical point extrapolations within different categories (i.e., characterised by variations of the input data as outlined in Subsection~\ref{ss:systematics}) outweighs the other one from the model averaging procedure itself (i.e., involving the various fit forms). Note that the statistical error of the correlation functions entering the individual definitions of the charm quark mass is by far subleading in our calculation and that the uncertainties on the experimental values of the meson masses, which feed into the analysis to set the physical point, are completely irrelevant for the total error of our central result (\ref{e:3f_rgi}).

As the final uncertainty of our charm quark mass determination is completely dominated by external input, there is hardly any room for a significant gain in precision, even if the underlying data base of gauge field configuration ensembles on the $\mathrm{Tr}[M_q]=\mathrm{const.}$ trajectory in lattice parameter space were moderately larger. One key prerequisite for an improved overall precision on the RGI (charm) quark mass along the methodology presented in this work would be a more accurate knowledge of the non-perturbative quark mass running factor. A promising step in that direction is the idea of a renormalisation strategy proposed in Ref.~\cite{DallaBrida:2019mqg} that exploits the decoupling of heavy quarks from low-energy physics. Another substantial refinement in precision may be anticipated from an update of the physical scale setting parameter, $\sqrt{8t_0^\mathrm{phys}}$, as it is currently being pursued within the CLS effort (see Refs.~\cite{Bali:2017bcv,BaliScale}).

The effects of charm loops are absent in our calculation, as we performed it on gauge ensembles with $2+1$ quark flavours in the sea. Still, for comparison to other results (from lattice and non-lattice QCD determinations) as well as later reference, it is customary to convert our result to the four-flavour theory through the well-established prescription relying on perturbative decoupling in the $\overline{\mathrm{MS}}$ scheme. This procedure is sketched in some detail in Appendix \ref{app:pt_mass_running}, where we also provide values of the charm quark mass in the $\overline{\mathrm{MS}}$ scheme at conventional energy scales.
Particularly, for the four-flavour RGI mass we arrive at:
\begin{align}
M_\mathrm{c}(N_\mathrm{f}=4, 5\text{-loop})&=1548(23)(2)_\Lambda\,\mathrm{MeV}=1548(23)\,\mathrm{MeV}\,.\label{e:4f_rgi}
\end{align}
Its error splits into the part derived from the uncertainty of the three-flavour RGI charm quark mass above, to which the subdominant part invoked by $\Lambda^{(3)}_{\overline{\mathrm{MS}}}=341(12)\,$MeV~\cite{Bruno:2017gxd} (that we employ for the QCD $\Lambda$-parameter entering the perturbative renormalisation group running) is added in quadrature. 

%% file: 06_conclusions.tex
\section{Summary and discussion}\label{s:discussion}
We have calculated the charm quark mass in QCD with $N_\mathrm{f}=2+1$ dynamical quark flavours, on the basis of a large set of gauge field configuration ensembles generated by the {\it Coordinated Lattice Simulations (CLS)} initiative \cite{Bruno:2014jqa,Mohler:2017wnb} and employing a partially quenched setup, consisting of non-perturbatively $\mathrm{O}(a)$ improved Wilson quarks in the sea supplemented by a quenched charm quark in the valence sector.
Bare current quark masses were extracted from the PCAC relation. Notable features of our determination are: five (very) fine lattice spacings from $0.087\,$fm down to less than $0.04\,$fm, fully non-perturbative $\mathrm{O}(a)$ improvement of the quark mass and the use of the non-perturbative renormalisation group running, available from Ref.~\cite{Campos:2018ahf}, in order to arrive at the RGI (renormalisation group invariant, i.e., scheme and scale independent) estimate of the charm quark mass as central result.
Moreover, our analysis exploits three different definitions of the quark mass on the lattice, which also allows systematic effects to be investigated in a controlled way.
To ensure reliability and robustness of our result that, for given possible variations in the analysis, includes a large enough (albeit theoretically well-founded) number of model functions for the joint chiral and continuum extrapolation whose individual contributions are correctly weighted by their fit to the actual data, we have adopted a model averaging procedure inspired by the prescription proposed in Ref.~\cite{Jay:2020jkz}. In this regard, our final result can therefore be considered as model-independent and its error as accounting for all systematic effects of the present computation in the studied $N_\mathrm{f}=2+1$ approximation of QCD.

The final numerical outcome of our work is to be taken from eqs.~(\ref{e:3f_rgi}) and (\ref{e:4f_rgi}) of the previous section, where we quote the RGI charm quark mass in the three-flavour (i.e., $N_\mathrm{f}=2+1$) and four-flavour theories, respectively, the latter being obtained by virtue of perturbative decoupling relations.
It is worth emphasising that the precision of our result can only be substantially improved by reducing the uncertainties on external quantities (most notably, the renormalised-to-RGI quark mass conversion factor and the lattice scale in physical units) which enter our calculation.
Furthermore, while the joint extrapolations of the charm quark mass to the physical point exhibit, in line with $\mathrm{O}(a)$ improvement, no corrections linearly, but quadratically and of higher order in the lattice spacing, they are not able to resolve any dependence on the sea pion mass; this stands in distinct contrast to the calculation of light quark masses on a subset of the gauge configuration ensembles at hand, reported in Ref.~\cite{Bruno:2019vup}, for which a decent decrease of errors can be expected from accommodating more chiral ensembles (especially with fine lattice spacings) and more statistics at light quark masses.

As soon as the relative error is smaller than about $1\%$, isospin splitting and QED effects may become noticeable. For instance, the impact of the inclusion of quenched QED was found to be around $0.2\%$ in Ref.~\cite{Hatton:2020qhk}, which is well below our final uncertainty. As apparent from Figure \ref{fig:flagcomparison} where we compare our result with other lattice QCD ones from the literature, no significant difference between $2+1$ and $2+1+1$ flavour results can be unravelled at the current level of precision of lattice QCD calculations.

After all, in Appendix \ref{app:pt_mass_running}, we explain how we translated eqs.~(\ref{e:3f_rgi}) and (\ref{e:4f_rgi}) to associated charm quark mass values in the $\overline{\mathrm{MS}}$ scheme, present our results at customary energy scales, and confront them with those of other lattice QCD collaborations.
Here, we compare our four-flavour RGI mass estimate $M_\mathrm{c}(N_\mathrm{f}=4)=1548(23)\,\mathrm{MeV}$ to the current FLAG average for calculations based on $N_\mathrm{f}=2+1$ dynamical quarks \cite{Aoki:2019cca}, which is composed of the numbers from Refs.~\cite{McNeile:2010ji,Yang:2014sea,Nakayama:2016atf} and reads:
\begin{align}
M_\mathrm{c}^\mathrm{FLAG}(N_\mathrm{f}=4)=1529(17)\,\mathrm{MeV}\,.
\end{align}
Within their $1\sigma$-errors, both are consistent with each other.

%% file: acknow.tex
We thank Mattia Bruno, Tomasz Korzec, Stefan Schaefer and Rainer Sommer, as well as Gunnar Bali, Sjoerd Bouma, Sara Collins and Wolfgang Söldner, for helpful discussions and the RQCD Collaboration for the fruitful collaboration in our joint project on charmed decay constants.
We are also indebted to Patrick Fritzsch, Carlos Pena and Anastassios Vladikas for a critical reading of an earlier version of the manuscript and their valuable comments.
We are grateful to our colleagues in the CLS initiative for producing the gauge field configuration ensembles used in this study. This work is supported by the Deutsche Forschungsgemeinschaft (DFG) through the Research Training Group {\it GRK 2149: Strong and Weak Interactions – from Hadrons to Dark Matter}. We acknowledge the computer resources provided by the WWU IT, formerly `Zentrum für Informationsverarbeitung (ZIV)', of the University of Münster (PALMA-II HPC cluster) and thank its staff for support.

%% file: app_quark_mass_running.tex
\section{Perturbative quark mass running to conventional scales}\label{a:running}
\label{app:pt_mass_running}
In view of further applications or reference in phenomenological studies, the charm quark mass is often quoted in the $\overline{\mathrm{MS}}$ renormalisation scheme and the four-flavour theory. In the case of heavy quarks, in particular, it is also common practice to do this in form of the scale invariant mass, which amounts to implicitly fixing the renormalisation (i.e., energy) scale of the running $\overline{\mathrm{MS}}$ quark mass at its own value.

For the purpose of this conversion, starting from the RGI mass value, we first use the result $\Lambda^{(3)}_{\overline{\mathrm{MS}}}=341(12)\,\mathrm{MeV}$ for the QCD $\Lambda$-parameter computed by the ALPHA Collaboration in Ref.~\cite{Bruno:2017gxd}, together with perturbative running in the $\overline{\mathrm{MS}}$ scheme and the three-flavour theory, to obtain the scale invariant charm quark mass, $m^{\overline{\mathrm{MS}}}_{\mathrm{c}}(\mu=m^{\overline{\mathrm{MS}}}_{\mathrm{c}}; N_\mathrm{f}=3)$; for the numerical evaluation, we proceed along the lines of Appendix A of Ref.~\cite{Bernardoni:2013xba}.
Next, at the energy scale defined by the scale invariant charm quark mass, one has to employ (perturbative) decoupling relations, which account for the effect of the change in the number of active quark flavours when passing across the respective quark mass thresholds while running with energy, in order to consistently arrive at a result in the four-flavour theory; for this step, we utilise the \texttt{RunDec} software package described in Refs.~\cite{Chetyrkin:2000yt,Schmidt:2012az,Herren:2017osy}. Then one can perturbatively run this number in the four-flavour theory, either to the scale invariant quark mass (which gets slightly shifted by the decoupling process) or upwards to the conventional energy scale $\mu=3\,$GeV. In this way we provide results, for both 4-loop and 5-loop perturbative accuracy of the $\beta$-function and the quark mass anomalous dimension,\footnote{Thanks to this high-order accuracy, perturbation theory that has to be invoked within any conversion to the conventionally cited $\overline{\mathrm{MS}}$ scheme looks very well behaved.}and find:
\begin{align}
m^{\overline{\mathrm{MS}}}_{\mathrm{c}}(\mu=m^{\overline{\mathrm{MS}}}_{\mathrm{c}}; N_\mathrm{f}=4, 4\text{-loop})&=1292(13)(14)_\Lambda\,\mathrm{MeV}=1292(19)\,\mathrm{MeV}\,,\\
m^{\overline{\mathrm{MS}}}_{\mathrm{c}}(\mu=m^{\overline{\mathrm{MS}}}_{\mathrm{c}}; N_\mathrm{f}=4, 5\text{-loop})&=1296(13)(14)_\Lambda\,\mathrm{MeV}=1296(19)\,\mathrm{MeV}\,;
\end{align}
and
\begin{align}
m^{\overline{\mathrm{MS}}}_{\mathrm{c}}(\mu=3\,\mathrm{GeV}; N_\mathrm{f}=4, 4\text{-loop})&=1005(14)(8)_\Lambda\,\mathrm{MeV}=1005(16)\,\mathrm{MeV}\,,\\
m^{\overline{\mathrm{MS}}}_{\mathrm{c}}(\mu=3\,\mathrm{GeV}; N_\mathrm{f}=4, 5\text{-loop})&=1007(14)(8)_\Lambda\,\mathrm{MeV}=1007(16)\,\mathrm{MeV}\,,
\end{align}
where the first error is the one from our determination discussed in the main text, whereas the second error stems from the uncertainty of $\Lambda^{(3)}_{\overline{\mathrm{MS}}}$.

After the switch to the four-flavour theory, we may employ perturbative running again in order to eventually obtain the four-flavour RGI charm quark mass, for which we quote
\begin{align}
M_\mathrm{c}(N_\mathrm{f}=4, 4\text{-loop})&=1548(23)(2)_\Lambda\,\mathrm{MeV}=1548(23)\,\mathrm{MeV}\,,
\end{align}
and the 5-loop value already stated in eq.~(\ref{e:4f_rgi}). Note that the error contribution from $\Lambda^{(3)}_{\overline{\mathrm{MS}}}$ becomes negligible in the four-flavour RGI limit and that the 4-loop and 5-loop results agree up to less than $1\,$MeV. We thus conclude that any ambiguities, which arise from the perturbative running and decoupling in the $\overline{\mathrm{MS}}$ scheme, are much smaller than the total error of our final result.

\begin{figure}[t]
	\centering
	\includegraphics[width=\linewidth]{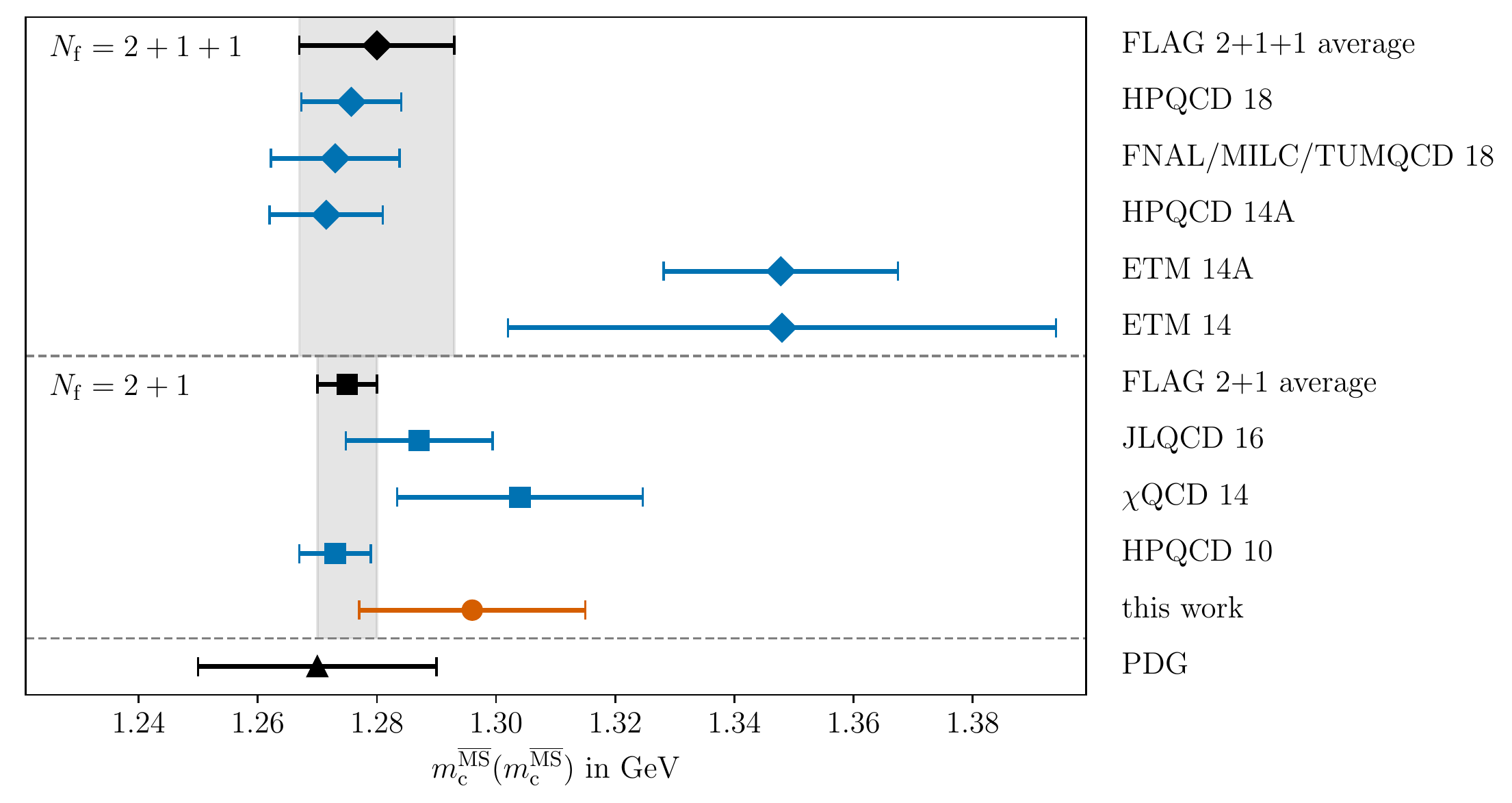}
	\caption{Comparison of our result (orange point) for $m^{\overline{\mathrm{MS}}}_{\mathrm{c}}(m^{\overline{\mathrm{MS}}}_{\mathrm{c}})$ in the four-flavour theory with $1.27(2)\,$GeV from the Particle Data Group \cite{Zyla:2020zbs} (black triangle), the $N_\mathrm{f}=2+1+1$ and $N_\mathrm{f}=2+1$ FLAG averages \cite{Aoki:2019cca} (grey vertical bands), as well as the lattice results that entered these averages: HPQCD 18 \cite{Lytle:2018evc}, FNAL/MILC/TUMQCD 18 \cite{Bazavov:2018omf}, HPQCD 14A \cite{Chakraborty:2014aca}, ETM 14A \cite{Alexandrou:2014sha}, ETM 14 \cite{Carrasco:2014cwa}, JLQCD 16 \cite{Nakayama:2016atf}, $\chi$QCD 14 \cite{Yang:2014sea}, HPQCD 10 \cite{McNeile:2010ji}. The upper part of the figure displays lattice results originating from simulations with four quark flavours in the sea (marked by diamonds), whereas the lower part refers to those obtained on gauge configurations with three dynamical quark flavours (marked by squares).}
	\label{fig:flagcomparison}
\end{figure}
In Figure~\ref{fig:flagcomparison} we compare our result for $m^{\overline{\mathrm{MS}}}_{\mathrm{c}}(m^{\overline{\mathrm{MS}}}_{\mathrm{c}})$ in the four-flavour theory with those from other lattice QCD determinations that entered the averages in the FLAG review in Ref.~\cite{Aoki:2019cca} based on gauge field configurations with $2+1$ and $2+1+1$ quark flavours in the sea. Except for the result from \cite{Alexandrou:2014sha}, labelled `ETM 14A' in Figure~\ref{fig:flagcomparison}, ours is compatible with all of the shown numbers within $1\sigma$ uncertainties. Given that the underlying strategies to extract the physical quark mass, as well as the discretised actions, renormalisation procedures and further potential systematics, differ significantly among these calculations, this overall agreement appears very reassuring. For completeness, we mention that our result is in line with the findings of Ref.~\cite{Hatton:2020qhk} for $2+1+1$ quark flavours and the result of Ref.~\cite{Petreczky:2019ozv} (which is an extension of Ref.~\cite{Maezawa:2016vgv}) for $2+1$ flavours. Both references have been published after the closing date for the FLAG review and do not show tension with the results in Figure~\ref{fig:flagcomparison}.

Let us close this appendix with a general remark on the renormalisation strategy which we follow in this work. Indeed, one might argue that it is not ideally suited when one is simply interested in the charm quark mass in the $\overline{\mathrm{MS}}$ scheme at a scale of a few GeV: namely, as a consequence of first running non-perturbatively (in the intermediate Schrödinger functional scheme) up to an energy scale of about $M_\mathrm{W}$ deeply in the perturbative regime, from which the scale evolution of the renormalised mass can safely be continued perturbatively to infinite energy where the RGI mass is defined, and then, again perturbatively, down to a few GeV in the $\overline{\mathrm{MS}}$ scheme, the errors from both running procedures dominate our uncertainty on $m^{\overline{\mathrm{MS}}}_{\mathrm{c}}(\mu=m^{\overline{\mathrm{MS}}}_{\mathrm{c}})$ by far.
Based on this observation, one could come up with the idea to only run non-perturbatively to the scale of choice, e.g., $2\,$GeV, and to perform the perturbative conversion to the $\overline{\mathrm{MS}}$ scheme directly at this scale. However, apart from the systematic uncertainties that are connected with the perturbative conversion at low-energy scales, the statistical error would not be decreased noticeably, since the dominant contribution to the uncertainty of the relevant factor encoding the non-perturbative running of the renormalised quark mass (analogous to (\ref{e:M_over_m_nf3})) is already accumulated in the low-energy regime, i.e., below $2\,$GeV.
Apart from these conceptual disadvantages of the $\overline{\mathrm{MS}}$ mass within our approach, we also prefer the RGI quark mass because of its a priori genuinely non-perturbative nature as well as its renormalisation scheme and scale independence and, therefore, distinctly more controllable systematic error.

%% file: app_tables.tex
\section{Tables} \label{app:result_tables}
\begin{table}[!htbp]
	\setlength{\abovecaptionskip}{10pt}
	\setlength{\belowcaptionskip}{0pt}
	\centering\small
	\renewcommand{\arraystretch}{1.25}
	\caption{Overview of the results for the gluonic quantity $t_0/a^2$ as well as the combinations $\phi_4$ and $\phi_2$ defined in eqs.~(\ref{e:phi_4}) and (\ref{e:phi_2}) from pion and kaon masses, together with the associated hopping parameters. For every gauge field configuration ensemble we list the numbers before the quark mass shift (see Section~\ref{s:quarkmasscomputation}) in the rows with labels `id' and the corresponding ones after the shift to $\phi_4^\mathrm{phys}$ in the rows without labels.}
	\input{./tables/tab_mesons.tex}
	\label{t:mesons}
\end{table}

\begin{table}[!htbp]
	\setlength{\abovecaptionskip}{10pt}
	\setlength{\belowcaptionskip}{0pt}
	\centering\small
	\renewcommand{\arraystretch}{1.25}
	\caption{Overview of the results for the flavour-averaged heavy-light and heavy-heavy pseudoscalar meson masses, referring to the two hopping parameter choices in the charm region given in Table \ref{t:kappa_h}. As before, for every ensemble we list the numbers before the shift in the rows with labels and the corresponding ones after the shift to $\phi_4^\mathrm{phys}$ in the rows without labels.}
	\input{./tables/tab_mesons_heavy.tex}
	\label{t:mesons_heavy}
\end{table}
\begin{table}[!htbp]
	\setlength{\abovecaptionskip}{10pt}
	\setlength{\belowcaptionskip}{0pt}
	\centering\small
	\renewcommand{\arraystretch}{1.25}
	\caption{Overview of the results for light-light, strange-strange and heavy-light PCAC quark masses from standard lattice derivatives, where the heavy quark hopping parameter choices in the charm region refer to Table \ref{t:kappa_h}. As before, for every ensemble we list the numbers before the shift in the rows with labels and the corresponding ones after the shift to $\phi_4^\mathrm{phys}$ in the rows without labels. Analogous results from improved lattice derivatives are collected in Table \ref{t:pcac_light_id}.}
	\input{./tables/tab_pcac_light_sd.tex}
	\label{t:pcac_light_sd}
\end{table}

\begin{table}[!htbp]
	\setlength{\abovecaptionskip}{10pt}
	\setlength{\belowcaptionskip}{0pt}
	\centering\small
	\renewcommand{\arraystretch}{1.25}
	\caption{Overview of the results for heavy-strange and heavy-heavy PCAC quark masses from standard lattice derivatives, referring to the heavy quark hopping parameter choices given in Table~\ref{t:kappa_h}. As before, for every ensemble we list the numbers before the shift in the rows with labels and the corresponding ones after the shift to $\phi_4^\mathrm{phys}$ in the rows without labels. Analogous results from improved lattice derivatives are collected in Table \ref{t:pcac_heavy_id}.}
	\input{./tables/tab_pcac_heavy_sd.tex}
	\label{t:pcac_heavy_sd}
\end{table}

\begin{table}[!htbp]
	\setlength{\abovecaptionskip}{10pt}
	\setlength{\belowcaptionskip}{0pt}
	\centering\small
	\renewcommand{\arraystretch}{1.25}
	\caption{Overview of the results for light-light, strange-strange and heavy-light PCAC quark masses from improved lattice derivatives, where the heavy quark hopping parameter choices in the charm region refer to Table \ref{t:kappa_h}. As before, for every ensemble we list the numbers before the shift in the rows with labels and the corresponding ones after the shift to $\phi_4^\mathrm{phys}$ in the rows without labels. Analogous results from standard lattice derivatives are collected in Table \ref{t:pcac_light_sd}.}
	\input{./tables/tab_pcac_light_id.tex}
	\label{t:pcac_light_id}
\end{table}

\begin{table}[!htbp]
	\setlength{\abovecaptionskip}{10pt}
	\setlength{\belowcaptionskip}{0pt}
	\centering\small
	\renewcommand{\arraystretch}{1.25}
	\caption{Overview of the results for heavy-strange and heavy-heavy PCAC quark masses from improved lattice derivatives, referring to the heavy quark hopping parameter choices given in Table \ref{t:kappa_h}. As before, for every ensemble we list the numbers before the shift in the rows with labels and the corresponding ones after the shift to $\phi_4^\mathrm{phys}$ in the rows without labels. Analogous results from standard lattice derivatives are collected in Table \ref{t:pcac_heavy_sd}.}
	\input{./tables/tab_pcac_heavy_id.tex}
	\label{t:pcac_heavy_id}
\end{table}

%% file: tables/tab_mesons.tex
\begin{tabular}{clllll}
\toprule
id & $t_0/a^2$ & $\phi_4$ & $\phi_2$ & $\kappa_\mathrm{l}$ & $\kappa_\mathrm{s}$ \\
\midrule
 H101 & $2.846( 8)$ & $1.141( 7)$ & $0.761( 5)$& 0.1367596200     & 0.1367596200     \\
      & $2.845( 8)$ & $1.120(25)$ & $0.747(17)$& 0.1367625(39) & 0.1367625(39) \\
 H102 & $2.872(12)$ & $1.112( 9)$ & $0.546( 6)$& 0.1368650000     & 0.1365493390     \\
      & $2.873(12)$ & $1.120(25)$ & $0.551(16)$& 0.1368641(30) & 0.1365485(30) \\
 H105 & $2.891(16)$ & $1.114(13)$ & $0.341( 8)$& 0.1369700000     & 0.1363407900     \\
      & $2.891(14)$ & $1.121(25)$ & $0.343(19)$& 0.1369693(32) & 0.1363401(32) \\
 C101 & $2.907( 7)$ & $1.100( 4)$ & $0.221( 3)$& 0.1370300000     & 0.1362220410     \\
      & $2.896(14)$ & $1.120(25)$ & $0.239(21)$& 0.1370269(33) & 0.1362190(33) \\
\midrule
 H400 & $3.634(13)$ & $1.163( 6)$ & $0.775( 4)$& 0.1368884800     & 0.1368884800     \\
      & $3.663(24)$ & $1.120(25)$ & $0.747(17)$& 0.1368982(65) & 0.1368982(65) \\
\midrule
 H200 & $5.149(29)$ & $1.151(10)$ & $0.768( 7)$& 0.1370000000     & 0.1370000000     \\
      & $5.170(35)$ & $1.120(25)$ & $0.747(17)$& 0.1370064(58) & 0.1370064(58) \\
 N202 & $5.140(21)$ & $1.111( 6)$ & $0.741( 4)$& 0.1370000000     & 0.1370000000     \\
      & $5.140(21)$ & $1.120(25)$ & $0.747(17)$& 0.1369983(39) & 0.1369983(39) \\
 N203 & $5.143( 7)$ & $1.112( 3)$ & $0.519( 3)$& 0.1370800000     & 0.1368402840     \\
      & $5.141(14)$ & $1.120(25)$ & $0.523(18)$& 0.1370791(41) & 0.1368394(41) \\
 N200 & $5.159( 8)$ & $1.114( 4)$ & $0.352( 2)$& 0.1371400000     & 0.1367208600     \\
      & $5.155(14)$ & $1.120(25)$ & $0.358(17)$& 0.1371387(38) & 0.1367195(38) \\
 D200 & $5.167(10)$ & $1.097( 3)$ & $0.173( 2)$& 0.1372000000     & 0.1366017480     \\
      & $5.167( 7)$ & $1.120(25)$ & $0.189(18)$& 0.1371964(35) & 0.1365982(35) \\
\midrule
 N300 & $8.564(24)$ & $1.151( 6)$ & $0.767( 4)$& 0.1370000000     & 0.1370000000     \\
      & $8.569(27)$ & $1.120(25)$ & $0.747(17)$& 0.1370057(48) & 0.1370057(48) \\
 N302 & $8.539(22)$ & $1.154(10)$ & $0.522( 4)$& 0.1370640000     & 0.1368721791     \\
      & $8.562(33)$ & $1.120(25)$ & $0.503(16)$& 0.1370687(43) & 0.1368769(43) \\
 J303 & $8.614(20)$ & $1.133( 5)$ & $0.290( 2)$& 0.1371230000     & 0.1367546608     \\
      & $8.612(17)$ & $1.120(25)$ & $0.282(15)$& 0.1371248(37) & 0.1367565(37) \\
\midrule
 J500 & $14.064(63)$ & $1.107(10)$ & $0.738( 7)$& 0.1368520000     & 0.1368520000     \\
      & $14.065(65)$ & $1.120(25)$ & $0.747(17)$& 0.1368509(27) & 0.1368509(27) \\
 J501 & $13.887(51)$ & $1.111( 8)$ & $0.491( 5)$& 0.1369032000     & 0.1367497150     \\
      & $13.883(59)$ & $1.120(25)$ & $0.494(17)$& 0.1369023(30) & 0.1367488(30) \\
\bottomrule
\end{tabular}

%% file: tables/tab_mesons_heavy.tex
\begin{tabular}{cllll}
\toprule
id & $am_{\bar{D}_1}$ & $am_{\bar{D}_2}$ & $am_{\eta_{\mathrm{c}, 1}}$ & $am_{\eta_{\mathrm{c}, 2}}$ \\
\midrule
 H101  & $0.8512(6)$ & $0.8203(5)$ & $1.3272(2)$ & $1.2726(2)$\\
       & $0.8508(12)$ & $0.8200(11)$ & $1.3269(4)$ & $1.2723(4)$\\
 H102  & $0.8500(6)$ & $0.8190(5)$ & $1.3260(3)$ & $1.2715(3)$\\
       & $0.8504(9)$ & $0.8194(9)$ & $1.3262(5)$ & $1.2716(5)$\\
 H105  & $0.8498(8)$ & $0.8188(8)$ & $1.3260(4)$ & $1.2714(4)$\\
       & $0.8500(9)$ & $0.8190(10)$ & $1.3261(5)$ & $1.2715(5)$\\
 C101  & $0.8494(5)$ & $0.8184(5)$ & $1.3253(2)$ & $1.2707(2)$\\
       & $0.8500(9)$ & $0.8190(9)$ & $1.3256(5)$ & $1.2710(5)$\\
\midrule
 H400  & $0.7450(6)$ & $0.6935(6)$ & $1.1614(3)$ & $1.0702(3)$\\
       & $0.7446(8)$ & $0.6931(8)$ & $1.1607(6)$ & $1.0694(6)$\\
\midrule
 H200  & $0.6470(7)$ & $0.6011(6)$ & $1.0115(2)$ & $0.9300(2)$\\
       & $0.6462(9)$ & $0.6003(10)$ & $1.0111(4)$ & $0.9296(4)$\\
 N202  & $0.6455(6)$ & $0.5996(6)$ & $1.0111(2)$ & $0.9295(2)$\\
       & $0.6457(6)$ & $0.5997(6)$ & $1.0112(3)$ & $0.9296(3)$\\
 N203  & $0.6438(7)$ & $0.5980(5)$ & $1.0112(1)$ & $0.9297(1)$\\
       & $0.6438(7)$ & $0.5979(7)$ & $1.0112(2)$ & $0.9297(2)$\\
 N200  & $0.6445(6)$ & $0.5986(5)$ & $1.0111(2)$ & $0.9295(2)$\\
       & $0.6444(7)$ & $0.5986(7)$ & $1.0111(3)$ & $0.9296(3)$\\
 D200  & $0.6435(5)$ & $0.5977(4)$ & $1.0110(1)$ & $0.9294(1)$\\
       & $0.6442(7)$ & $0.5983(6)$ & $1.0111(2)$ & $0.9295(2)$\\
\midrule
 N300  & $0.4929(4)$ & $0.4751(4)$ & $0.7731(2)$ & $0.7412(2)$\\
       & $0.4927(5)$ & $0.4749(5)$ & $0.7731(2)$ & $0.7411(2)$\\
 N302  & $0.4929(5)$ & $0.4751(5)$ & $0.7732(2)$ & $0.7413(2)$\\
       & $0.4923(7)$ & $0.4745(9)$ & $0.7730(2)$ & $0.7411(2)$\\
 J303  & $0.4926(4)$ & $0.4748(4)$ & $0.7728(1)$ & $0.7409(1)$\\
       & $0.4925(5)$ & $0.4746(5)$ & $0.7727(2)$ & $0.7408(2)$\\
\midrule
 J500  & $0.3712(8)$ & $0.3553(8)$ & $0.5814(2)$ & $0.5529(2)$\\
       & $0.3713(8)$ & $0.3554(8)$ & $0.5814(2)$ & $0.5530(2)$\\
 J501  & $0.3723(5)$ & $0.3563(5)$ & $0.5814(2)$ & $0.5530(2)$\\
       & $0.3723(5)$ & $0.3564(5)$ & $0.5814(2)$ & $0.5530(2)$\\
\bottomrule
\end{tabular}

%% file: tables/tab_pcac_light_sd.tex
\begin{tabular}{cllll}
\toprule
id & $am_{\mathrm{ll^\prime}}$  & $am_{\mathrm{ss^\prime}}$ & $am_{\mathrm{h}_1\mathrm{l}}$ & $am_{\mathrm{h}_2\mathrm{l}}$\\
\midrule
 H101  & $0.009207(46)$ & $0.009207(46)$ & $0.190403(67)$ & $0.177377(61)$\\
       & $0.00905(19)$ & $0.00905(19)$ & $0.19025(22)$ & $0.17725(20)$\\
 H102  & $0.006508(48)$ & $0.013828(51)$ & $0.188412(79)$ & $0.175403(68)$\\
       & $0.00658(18)$ & $0.01390(19)$ & $0.18850(16)$ & $0.17549(16)$\\
 H105  & $0.003900(83)$ & $0.018612(86)$ & $0.186583(186)$ & $0.173604(171)$\\
       & $0.00393(20)$ & $0.01866(19)$ & $0.18661(23)$ & $0.17365(22)$\\
 C101  & $0.002502(38)$ & $0.021265(68)$ & $0.185763(98)$ & $0.172776(84)$\\
       & $0.00271(25)$ & $0.02144(24)$ & $0.18593(21)$ & $0.17294(20)$\\
\midrule
 H400  & $0.008232(37)$ & $0.008232(37)$ & $0.160086(72)$ & $0.140195(78)$\\
       & $0.00785(22)$ & $0.00785(22)$ & $0.15979(22)$ & $0.13989(23)$\\
\midrule
 H200  & $0.006860(26)$ & $0.006860(26)$ & $0.137317(51)$ & $0.120424(36)$\\
       & $0.00664(19)$ & $0.00664(19)$ & $0.13717(17)$ & $0.12027(17)$\\
 N202  & $0.006849(18)$ & $0.006849(18)$ & $0.137281(48)$ & $0.120382(37)$\\
       & $0.00692(16)$ & $0.00692(16)$ & $0.13734(12)$ & $0.12043(11)$\\
 N203  & $0.004745(18)$ & $0.011047(12)$ & $0.135989(43)$ & $0.119136(32)$\\
       & $0.00478(18)$ & $0.01108(17)$ & $0.13602(11)$ & $0.11916(11)$\\
 N200  & $0.003152(14)$ & $0.014151(15)$ & $0.135006(37)$ & $0.118165(37)$\\
       & $0.00321(17)$ & $0.01421(17)$ & $0.13506(12)$ & $0.11822(12)$\\
 D200  & $0.001530(12)$ & $0.017239(10)$ & $0.134032(43)$ & $0.117212(35)$\\
       & $0.00169(18)$ & $0.01739(16)$ & $0.13411(9)$ & $0.11730(10)$\\
\midrule
 N300  & $0.005507(7)$ & $0.005507(7)$ & $0.100574(12)$ & $0.094440(11)$\\
       & $0.00534(14)$ & $0.00534(14)$ & $0.10048(9)$ & $0.09434(9)$\\
 N302  & $0.003722(13)$ & $0.009086(11)$ & $0.099580(22)$ & $0.093451(18)$\\
       & $0.00357(13)$ & $0.00893(14)$ & $0.09949(10)$ & $0.09336(10)$\\
 J303  & $0.002051(8)$ & $0.012352(7)$ & $0.098653(20)$ & $0.092529(18)$\\
       & $0.00199(11)$ & $0.01229(12)$ & $0.09863(8)$ & $0.09250(8)$\\
\midrule
 J500  & $0.004211(5)$ & $0.004211(5)$ & $0.072220(10)$ & $0.067129(9)$\\
       & $0.00425(10)$ & $0.00425(10)$ & $0.07225(6)$ & $0.06716(6)$\\
 J501  & $0.002740(6)$ & $0.007173(5)$ & $0.071461(13)$ & $0.066370(11)$\\
       & $0.00277(11)$ & $0.00720(11)$ & $0.07148(6)$ & $0.06639(6)$\\
\bottomrule
\end{tabular}

%% file: tables/tab_pcac_heavy_sd.tex
\begin{tabular}{cllllll}
\toprule
id & $am_{\mathrm{h}_1\mathrm{s}}$ & $am_{\mathrm{h}_2\mathrm{s}}$ & $am_{\mathrm{h}_1\mathrm{h}^\prime_1}$ & $am_{\mathrm{h}_2\mathrm{h}^\prime_2}$\\
\midrule
 H101  & $0.190403(67)$ & $0.177377(61)$ & $0.443767(70)$ & $0.407199(65)$\\
       & $0.19025(22)$ & $0.17725(20)$ & $0.44365(16)$ & $0.40708(15)$\\
 H102  & $0.193532(61)$ & $0.180437(59)$ & $0.443399(76)$ & $0.406843(71)$\\
       & $0.19360(17)$ & $0.18050(17)$ & $0.44346(15)$ & $0.40689(15)$\\
 H105  & $0.196843(95)$ & $0.183687(89)$ & $0.443335(120)$ & $0.406780(115)$\\
       & $0.19690(17)$ & $0.18374(17)$ & $0.44337(16)$ & $0.40682(15)$\\
 C101  & $0.198621(45)$ & $0.185421(44)$ & $0.443063(67)$ & $0.406520(66)$\\
       & $0.19882(18)$ & $0.18561(19)$ & $0.44321(17)$ & $0.40666(16)$\\
\midrule
 H400  & $0.160086(72)$ & $0.140195(78)$ & $0.357665(65)$ & $0.305959(61)$\\
       & $0.15979(22)$ & $0.13989(23)$ & $0.35742(17)$ & $0.30572(16)$\\
\midrule
 H200  & $0.137317(51)$ & $0.120424(36)$ & $0.296844(28)$ & $0.255337(26)$\\
       & $0.13717(17)$ & $0.12027(17)$ & $0.29676(9)$ & $0.25526(8)$\\
 N202  & $0.137281(48)$ & $0.120382(37)$ & $0.296801(25)$ & $0.255298(22)$\\
       & $0.13734(12)$ & $0.12043(11)$ & $0.29683(6)$ & $0.25532(6)$\\
 N203  & $0.139776(34)$ & $0.122822(25)$ & $0.296815(13)$ & $0.255310(12)$\\
       & $0.13979(11)$ & $0.12285(10)$ & $0.29683(6)$ & $0.25532(6)$\\
 N200  & $0.141613(25)$ & $0.124620(20)$ & $0.296765(20)$ & $0.255261(16)$\\
       & $0.14165(12)$ & $0.12466(11)$ & $0.29679(6)$ & $0.25528(6)$\\
 D200  & $0.143513(16)$ & $0.126461(13)$ & $0.296725(12)$ & $0.255219(10)$\\
       & $0.14362(11)$ & $0.12657(10)$ & $0.29677(5)$ & $0.25526(5)$\\
\midrule
 N300  & $0.100574(12)$ & $0.094440(11)$ & $0.207658(9)$ & $0.193678(7)$\\
       & $0.10048(9)$ & $0.09434(9)$ & $0.20763(3)$ & $0.19365(3)$\\
 N302  & $0.102546(19)$ & $0.096398(18)$ & $0.207660(10)$ & $0.193677(9)$\\
       & $0.10246(10)$ & $0.09631(10)$ & $0.20763(4)$ & $0.19364(4)$\\
 J303  & $0.104351(11)$ & $0.098186(10)$ & $0.207619(6)$ & $0.193639(5)$\\
       & $0.10431(9)$ & $0.09815(9)$ & $0.20761(3)$ & $0.19363(3)$\\
\midrule
 J500  & $0.072220(10)$ & $0.067129(9)$ & $0.145009(6)$ & $0.134031(5)$\\
       & $0.07225(6)$ & $0.06716(6)$ & $0.14501(2)$ & $0.13404(2)$\\
 J501  & $0.073800(9)$ & $0.068699(7)$ & $0.145012(5)$ & $0.134036(4)$\\
       & $0.07382(7)$ & $0.06872(6)$ & $0.14502(2)$ & $0.13404(2)$\\
\bottomrule
\end{tabular}

%% file: tables/tab_pcac_light_id.tex
\begin{tabular}{cllll}
\toprule
id & $am_{\mathrm{ll^\prime}}$  & $am_{\mathrm{ss^\prime}}$ & $am_{\mathrm{h}_1\mathrm{l}}$ & $am_{\mathrm{h}_2\mathrm{l}}$\\
\midrule
 H101  & $0.009153(45)$ & $0.009153(45)$ & $0.164694(49)$ & $0.155237(49)$\\
       & $0.00900(19)$ & $0.00900(19)$ & $0.16460(13)$ & $0.15515(13)$\\
 H102  & $0.006466(47)$ & $0.013705(49)$ & $0.163359(40)$ & $0.153873(38)$\\
       & $0.00654(19)$ & $0.01377(18)$ & $0.16337(9)$ & $0.15390(10)$\\
 H105  & $0.003896(86)$ & $0.018392(83)$ & $0.162020(77)$ & $0.152530(72)$\\
       & $0.00393(19)$ & $0.01844(19)$ & $0.16204(12)$ & $0.15255(11)$\\
 C101  & $0.002498(37)$ & $0.020984(64)$ & $0.161379(55)$ & $0.151853(55)$\\
       & $0.00270(25)$ & $0.02116(23)$ & $0.16147(13)$ & $0.15196(13)$\\
\midrule
 H400  & $0.008190(36)$ & $0.008190(36)$ & $0.143921(58)$ & $0.128051(54)$\\
       & $0.00781(22)$ & $0.00781(22)$ & $0.14368(17)$ & $0.12777(19)$\\
\midrule
 H200  & $0.006835(25)$ & $0.006835(25)$ & $0.127114(32)$ & $0.112719(44)$\\
       & $0.00661(19)$ & $0.00661(19)$ & $0.12697(14)$ & $0.11258(14)$\\
 N202  & $0.006826(17)$ & $0.006826(17)$ & $0.127129(27)$ & $0.112733(38)$\\
       & $0.00690(16)$ & $0.00690(16)$ & $0.12717(10)$ & $0.11278(10)$\\
 N203  & $0.004735(17)$ & $0.010993(12)$ & $0.126094(28)$ & $0.111740(32)$\\
       & $0.00477(17)$ & $0.01102(17)$ & $0.12611(8)$ & $0.11176(8)$\\
 N200  & $0.003147(14)$ & $0.014061(15)$ & $0.125244(34)$ & $0.110847(56)$\\
       & $0.00320(17)$ & $0.01412(16)$ & $0.12528(11)$ & $0.11088(11)$\\
 D200  & $0.001529(12)$ & $0.017106(10)$ & $0.124424(33)$ & $0.110066(83)$\\
       & $0.00169(18)$ & $0.01725(15)$ & $0.12448(7)$ & $0.11010(9)$\\
\midrule
 N300  & $0.005495(6)$ & $0.005495(6)$ & $0.096331(21)$ & $0.090717(6)$\\
       & $0.00533(14)$ & $0.00533(14)$ & $0.09621(9)$ & $0.09063(9)$\\
 N302  & $0.003717(13)$ & $0.009057(11)$ & $0.095426(9)$ & $0.089837(10)$\\
       & $0.00357(13)$ & $0.00890(14)$ & $0.09534(9)$ & $0.08975(9)$\\
 J303  & $0.002049(8)$ & $0.012301(6)$ & $0.094594(16)$ & $0.089002(15)$\\
       & $0.00199(11)$ & $0.01224(12)$ & $0.09457(7)$ & $0.08898(7)$\\
\midrule
 J500  & $0.004207(4)$ & $0.004207(4)$ & $0.070512(8)$ & $0.065673(8)$\\
       & $0.00424(10)$ & $0.00424(10)$ & $0.07053(5)$ & $0.06570(5)$\\
 J501  & $0.002737(6)$ & $0.007158(5)$ & $0.069775(7)$ & $0.064936(7)$\\
       & $0.00277(11)$ & $0.00719(11)$ & $0.06980(6)$ & $0.06496(6)$\\
\bottomrule
\end{tabular}

%% file: tables/tab_pcac_heavy_id.tex
\begin{tabular}{cllllll}
\toprule
id & $am_{\mathrm{h}_1\mathrm{s}}$ & $am_{\mathrm{h}_2\mathrm{s}}$ & $am_{\mathrm{h}_1\mathrm{h}^\prime_1}$ & $am_{\mathrm{h}_2\mathrm{h}^\prime_2}$\\
\midrule
 H101  & $0.164694(49)$ & $0.155237(49)$ & $0.284757(19)$ & $0.274638(14)$\\
       & $0.16460(13)$ & $0.15515(13)$ & $0.28476(2)$ & $0.27463(2)$\\
 H102  & $0.167030(33)$ & $0.157584(32)$ & $0.284822(20)$ & $0.274671(13)$\\
       & $0.16706(10)$ & $0.15761(10)$ & $0.28481(3)$ & $0.27467(2)$\\
 H105  & $0.169323(44)$ & $0.159911(46)$ & $0.284795(26)$ & $0.274642(15)$\\
       & $0.16934(10)$ & $0.15993(10)$ & $0.28479(3)$ & $0.27464(2)$\\
 C101  & $0.170660(26)$ & $0.161242(27)$ & $0.284842(16)$ & $0.274665(10)$\\
       & $0.17074(10)$ & $0.16133(11)$ & $0.28483(3)$ & $0.27466(2)$\\
\midrule
 H400  & $0.143921(58)$ & $0.128051(54)$ & $0.263692(14)$ & $0.238888(14)$\\
       & $0.14368(17)$ & $0.12777(19)$ & $0.26365(3)$ & $0.23882(5)$\\
\midrule
 H200  & $0.127114(32)$ & $0.112719(44)$ & $0.239900(10)$ & $0.214508(11)$\\
       & $0.12697(14)$ & $0.11258(14)$ & $0.23987(3)$ & $0.21447(4)$\\
 N202  & $0.127129(27)$ & $0.112733(38)$ & $0.239909(6)$ & $0.214510(6)$\\
       & $0.12717(10)$ & $0.11278(10)$ & $0.23991(2)$ & $0.21452(3)$\\
 N203  & $0.129254(22)$ & $0.114886(21)$ & $0.239909(6)$ & $0.214508(5)$\\
       & $0.12927(8)$ & $0.11490(8)$ & $0.23991(2)$ & $0.21451(3)$\\
 N200  & $0.130767(18)$ & $0.116393(16)$ & $0.239879(6)$ & $0.214481(6)$\\
       & $0.13080(9)$ & $0.11642(9)$ & $0.23988(2)$ & $0.21449(3)$\\
 D200  & $0.132339(11)$ & $0.117978(13)$ & $0.239865(4)$ & $0.214461(4)$\\
       & $0.13242(8)$ & $0.11806(9)$ & $0.23988(2)$ & $0.21448(2)$\\
\midrule
 N300  & $0.096331(21)$ & $0.090717(6)$ & $0.185473(4)$ & $0.174740(4)$\\
       & $0.09621(9)$ & $0.09063(9)$ & $0.18545(2)$ & $0.17471(3)$\\
 N302  & $0.098118(8)$ & $0.092529(12)$ & $0.185467(5)$ & $0.174735(4)$\\
       & $0.09804(8)$ & $0.09245(9)$ & $0.18545(3)$ & $0.17471(3)$\\
 J303  & $0.099755(8)$ & $0.094167(8)$ & $0.185456(3)$ & $0.174722(3)$\\
       & $0.09973(7)$ & $0.09414(7)$ & $0.18545(2)$ & $0.17472(2)$\\
\midrule
 J500  & $0.070512(8)$ & $0.065673(8)$ & $0.136498(3)$ & $0.126938(3)$\\
       & $0.07053(5)$ & $0.06570(5)$ & $0.13650(1)$ & $0.12694(1)$\\
 J501  & $0.072000(5)$ & $0.067162(4)$ & $0.136501(3)$ & $0.126939(3)$\\
       & $0.07202(6)$ & $0.06718(6)$ & $0.13650(1)$ & $0.12694(1)$\\
\bottomrule
\end{tabular}